\documentclass[a4paper,11pt]{article}
\pdfoutput=1

\usepackage{jheppub}

\usepackage[makeroom]{cancel}
\usepackage{graphicx}
\usepackage{hyperref}
\usepackage{orcidlink}
\usepackage{multirow}
\usepackage{overpic}

\def \bit{\begin{itemize}}
\def \eit{\end{itemize}}
\def \bea{\begin{eqnarray}}
\def \eea{\end{eqnarray}}
\def \beq{\begin{equation}}
\def \eeq{\end{equation}}
\def \[{\left[}
\def \]{\right]}
\def \({\left(}
\def \){\right)}
\def \lb{\left\{}
\def \rb{\right\}}

\def \bqd{{\bar q_d}}
\def \bl{{\bar\ell}}
\def \bL{{\bar L}}

\def \nn{\nonumber}
\def \nl{\nn\\}
\def \cB{{\cal B}}
\def \cL{{\cal L}}
\def \cH{{\cal H}}
\def \Re{{\rm Re}}
\def \Im{{\rm Im}}
\def \hc{\,+~{\rm h.c.}}
\def \eff{{\rm eff}}

\def\bctaunutau {b\to c\tau^-{\bar\nu}_\tau}

\def \al{\alpha}
\def \be{\beta}
\def \ga{\gamma}
\def \Ga{\Gamma}

\def \si{\sigma}

\title{Neutrino Nonstandard Interactions and  Lepton Flavor Universality violation at SND@LHC via charm production}

\author[a]{Bhubanjyoti Bhattacharya\,\orcidlink{0000-0003-2238-321X},}
\author[b]{Alakabha Datta\,\orcidlink{0000-0001-8713-2783},}
\author[c,d]{Elena Graverini\,\orcidlink{0000-0003-4647-6429},}
\author[e]{Lopamudra Mukherjee\,\orcidlink{0000-0001-8765-7563},}
\author[f,g]{Divya Sachdeva,}
\author[h]{and John Waite}

\affiliation[a]{Department of Natural Sciences, Lawrence Technological University, Southfield, MI 48075, USA}
\affiliation[b]{Department of Physics and Astronomy, 108 Lewis Hall, University of Mississippi, Oxford, MS38677-1848, USA}
\affiliation[c]{Università di Pisa, Italy}
\affiliation[d]{École Polytechnique Fédérale, Lausanne, Switzerland}
\affiliation[e]{School of Physics, Nankai University, Tianjin 300071, China}
\affiliation[f]{Laboratoire de Physique de l'\'Ecole normale sup\'erieure, ENS, Universit\'e PSL, CNRS, Sorbonne Universit\'e, Universit\'e Paris Cit\'e, 24 Rue Lhomond, F-75005 Paris, France}
\affiliation[g]{Indian Institute of Technology Hyderabad, Kandi, Sangareddy-502285, Telangana, India}
\affiliation[h]{Morehead State University, Morehead, KY, USA}

\emailAdd{bbhattach@ltu.edu}
\emailAdd{datta@phy.olemiss.edu}
\emailAdd{elena.graverini@cern.ch}
\emailAdd{lopamudra.physics@gmail.com}
\emailAdd{divyasachdeva@phy.iith.ac.in}
\emailAdd{j.waite@moreheadstate.edu}

\abstract{
In this work, we explore the effect of neutrino nonstandard interactions (NSI) involving the charm quark at SND@LHC. Using an effective description of new physics in terms of four-fermion operators involving a charm quark, we constrain the Wilson coefficients of the effective interaction from two and three-body charmed meson decays. In our fit, we include charmed meson decays not only to pseudoscalar final states but also to vector final states and include decays to the $\eta$ and $\eta^\prime$ final states. We also consider constraints from charmed baryon decays. We then study the effect of new physics in neutrino scattering processes, involving charm production at SND@LHC, for various benchmark new physics couplings obtained from the low energy fits. Finally, we also study the effects of lepton universality violation (LUV) assuming that the new physics coupling is not lepton universal.}

\begin{document}
\maketitle
\flushbottom

\section{Introduction} 

The observation of neutrino masses and mixing indicate physics beyond the Standard Model (SM). Therefore, it is reasonable to hypothesize that neutrinos may have new interactions beyond the SM. These interactions, known as neutrino nonstandard interactions (NSI), can be explored in specific models or in a model-independent framework in terms of four-fermion operators. The interactions may be purely leptonic or semileptonic where in the latter case a quark current and a leptonic current are involved in the effective interaction. In typical neutrino experiments, NSI involving the first-generation quarks are involved. In many models of BSM physics, the new physics (NP) effects are more pronounced in the heavier generations. NSI involving the heavy quarks and leptons are particularly interesting in these models. Also, given that there are hints of NP in decays of heavy quarks, a program to explore NSI involving heavy quarks is quite compelling. New high-energy neutrino scattering experiments that will produce charm quarks offer a unique opportunity to study NSI with heavy quarks.

The SM is lepton-flavor universal, i.e. the electroweak gauge interactions of the SM apply identically to all three flavors of leptons. Violation of this universality, known as lepton universality violation (LUV), is a crucial test of the SM. Evidence of LUV in $B$-mesons decays has generated significant interest in testing for LUV in various decays. In $B$ decays, hints of LUV have been observed in the charged-current quark-level transition $b\to c\ell^-\bar\nu$, where $\ell$ can be an $e, \mu,$ or $\tau$. At the mesonic level, since the hadronic part of such a decay rate is plagued by uncertainties from form factors and the non-perturbative nature of QCD, one considers ratios of decay rates as relatively cleaner tests for LUV. 

The SM predictions for the ratios $R^{\tau/\ell}_{D^{(*)}} \equiv \cB({\bar B}\to D^{(*)}\tau^-{\bar\nu_\tau})/\cB({\bar B}\to D^{(*)}\ell^-{\bar\nu_\ell})$ (here $\ell = e,\mu$) and  $R^{\tau/\mu}_{J/\psi} \equiv \cB(B_c^+\to J/\psi\tau^+\nu_\tau)/\cB(B_c^+\to J/\psi\mu^+\nu_\mu)$ are known to within 2\% and 20\% respectively \cite{HFLAV:2022esi}. The BaBar, Belle, and LHCb experiments have measured $R^{\tau/\ell}_{D^{(*)}}$ \cite{Lees:2012xj, Lees:2013uzd, Aaij:2015yra, Huschle:2015rga, Sato:2016svk, Hirose:2016wfn, Aaij:2017uff, Hirose:2017dxl, Aaij:2017deq, Belle:2019gij, LHCb:2023zxo, LHCb:2023uiv, Belle-II:2024ami, LHCb:2024jll} and $R^{\tau/\mu}_{J/\psi}$ \cite{Aaij:2017tyk} with a precision of 5-8\% and 35\% respectively. These measurements currently display some tension with their corresponding SM expectations. In particular, the combined deviation from the SM in $R^{\tau/\ell}_{D^{(*)}}$ is 3.31$\sigma$ \cite{HFLAV:RDRDst2024update}, while in $R^{\tau/\mu}_{J/\psi}$ the deviation is 1.7$\sigma$. 
Together these measurements provide strong hints of LUV NP in $\bctaunutau$ decays.

Current experimental measurements also allow a 3-5\% LUV in the ratio $R_{D^{(*)}}\equiv \cB({\bar B}\to D^{(*)} \mu^-{\bar\nu}_\mu)/\cB({\bar B}\to D^{(*)}e^-{\bar\nu}_e)$, which is expected to be $\sim 1$ in the SM. For the interested reader, we summarize the SM predictions for these observables and their experimental measurements in Appendix \ref{app:LUVr}.

Hints of NP are not restricted to charged-current $B$ decays. A recent first measurement of the branching ratio $\cB(B^+\to K^+\nu\bar\nu) = (2.3\pm 0.7)\times 10^{-5}$ by the Belle II experiment~\cite{Belle-II:2023esi} is $2.7\sigma$ higher than the SM expectation
$\cB(B^+\to K^+ \nu\bar\nu)_{\rm SM}=(5.58\pm 0.38)\times 10^{-6}$~\cite{Parrott:2022zte}. Furthermore, even though the recently updated measurements of $R_{K^{(*)}} = \cB(B\to K^{(*)} \mu^+\mu^-)/\cB(B\to K^{(*)} e^+ e^-)$ are now fully consistent with their SM expectations~\cite{LHCb:2022qnv}, the individual branching fractions in both the electron and muon channels remain discrepant~\cite{Capdevila:2023yhq}). Joint explanations of all these anomalies, both in model-dependent as well as model-independent approaches, favor NP that affects the heavier generations of quarks and leptons (see for example Refs.~\cite{Bhattacharya:2016mcc, Bhattacharya:2014wla}.)

While many of these anomalies have been observed in low-energy data, if NP is involved, it should also affect quark-lepton interactions at higher energy scales. In this paper, we study the effects of NP, specifically in the neutrino-quark scattering process $\nu_\ell + q \to q' + \ell^-$ where $\ell$ can be $e, \mu,$ or $\tau$. The effects of LUV NP in the scattering of $\nu_\tau$ off light quarks were studied in Refs.~\cite{Rashed:2012bd, Rashed:2013dba, Liu:2015rqa}. The parameter space of NP couplings and energy scales for $\nu_\tau$ scattering off light quarks can be constrained using data from hadronic tau decays. However, since the light leptons do not decay hadronically, similar constraints do not appear for $\nu_e$ and $\nu_\mu$. Instead, if the scattering of $\nu_\ell$ off of a light quark ($q$) produces a heavy quark ($h$) then the corresponding parameter space can be constrained by studying semileptonic transitions of a heavy quark to a light quark, $h\to q\ell^-{\bar\nu}$.  

Two recently started experiments that can facilitate the exploration of neutrino-quark scattering processes are FASER$\nu$~\cite{FASER:2019dxq} and the Scattering and Neutrino Detector at the Large Hadron Collider (SND@LHC)~\cite{SNDLHC:2022ihg}. Both detectors have been designed to investigate the scattering of high-energy neutrinos produced in the far-forward direction at the LHC, specifically at the ATLAS interaction point. In this work, we focus on the physics reach of the SND@LHC experiment. A recent analysis of NSI at FASER$\nu$ can be found in Ref.~\cite{Falkowski:2021bkq}.

Neutrinos reaching the SND@LHC detector are energetic enough to produce heavy quarks, such as the charm quark, in the final state. This allows SND@LHC to be a laboratory for testing LUV effects in quark-neutrino scattering processes involving a heavy quark. The SND@LHC detector also aims at exploring the possibility of detecting new particles that scatter similar to neutrinos, such as light dark matter (LDM) particles which interact with SM particles through portal mediators. This study focuses on estimating the sensitivity of the SND@LHC detector in detecting new-physics phenomena, considering various benchmark new-physics couplings derived from low-energy fits.

The SND@LHC detector comprises a target region followed by the muon system. The detector pseudorapidity range spans from 7.2 to 8.4. The target consists of five walls of emulsion cloud chambers (ECC) followed by planes of Scintillating Fiber (SciFi) trackers. Each wall comprises 60 emulsion films interleaved with 59 tungsten plates, each 1 mm thick, serving as the target material. The ECC provides micrometric accuracy for measuring charged-particle tracks and reconstructing vertices of neutrino interactions. The reconstruction of particles and showers spanning several emulsion bricks is aided by the interleaved layers of SciFi, that provide accurate time stamps. When combined with the ECC walls used as radiators, the SciFi detector also serves as a sampling calorimeter to measure the energy of electromagnetic showers.
Hadronic showers start developing in the target volume, but then they are fully contained by a hadronic calorimeter composed of alternating layers of 20 cm thick iron walls and 1 cm thick scintillating bars.

A key feature of SND@LHC is its high efficiency in identifying neutrino flavors. Electron neutrinos can be identified via their electromagnetic showers with 99\% efficiency. Charged-current muon neutrino interactions can be detected with 69\% efficiency by requiring the presence of at least one muon track. Neutral current interactions are correctly tagged with 99\% efficiency. Thanks to the micrometric resolution of the ECC, tau neutrinos can be identified via a decay vertex that is displaced from the primary interaction point. The $\nu_\tau$ detection efficiency for SND@LHC ranges from 48\% to 54\% depending on the decay mode of the $\tau$ lepton~\cite{Ahdida:2750060, SNDLHC:2022ihg}. 

In this work, we will explore LUV effects in the scattering process $\nu_\ell + q \to c + \ell^-$ at SND@LHC, where $q$ represents down-type quarks in the target (tungsten) nuclei. The incoming neutrinos are highly energetic so deep-inelastic scattering (DIS) is the underlying process. In the SM, the contributions from a $b$-quark in the initial state are suppressed due to the relative smallness of the Cabibbo-Kobayashi-Maskawa (CKM) matrix element $V_{cb}$ compared to contributions from a $d$ or $s$ quark in the initial state. For the same reason, the scattering process where an up-type quark from the target nuclei contributes in the initial state to produce a $b$ quark in the final state is also suppressed and is not considered here. We adopt an effective field theory (EFT) framework starting with all dimension 6 operators that can contribute to the process. We then constrain the Wilson Coefficients (WCs) of the effective operators using low-energy measurements of two- and three-body $D$ decays. While constraints on NP from semileptonic charmed mesons have been considered before \cite{Becirevic:2020rzi}, here we present a more exhaustive fit to the low-energy charm semileptonic decays. In our low-energy observables, we consider both two-body leptonic and three-body semileptonic decays of charmed mesons where we include decays not only to pseudoscalar final states but also to vector final states and include in our fits decays to the $\eta$ and $\eta'$ mesons. We also consider constraints from charmed baryon decays.
After the low energy fit, various benchmark values for the WCs are then used to study the prospects of detecting NP at SND@LHC. Although our results are specifically derived for the case of SND@LHC, they can be easily adapted and rescaled for application to FASER$\nu$.

The paper is organized in the following manner. In section II, we describe the formalism of the effective Hamiltonian and discuss several UV complete models from which the effective operators in the Hamiltonian can emerge. In section III, we discuss the low-energy flavor observables, perform fits to constrain the WCs, and choose benchmark values for the neutrino scattering study. In section IV, we consider the neutrino scattering and discuss the level of significance at which LUV effects can be detected at SND@LHC. We summarize and present our conclusions in section V.

\section{Effective Hamiltonian}

We begin by considering an effective Hamiltonian with dimension 6 four-fermion operators and Wilson coefficients that encode high-energy NP. Using the effective Hamiltonian we choose benchmark values for Wilson coefficients using constraints from low-energy observables including leptonic and semileptonic decays of $D$ mesons. The benchmark values are then used to study the effects of NP on neutrino scattering. The effective Hamiltonian that contributes to the scattering process $\nu_\ell + q \to c + \ell^-$ can be expressed as \cite{Falkowski:2021bkq,Kopp:2024yvh}:
\bea
\label{eq:Heffo}
H_{\rm eff} &=& \frac{4G_FV_{cq}}{\sqrt{2}}\left\{(1+\epsilon_L)_{\al\be}[{\bar c}\ga_\mu P_L q][{\bl}_\al\ga^\mu P_L\nu_\be] + (\epsilon_R)_{\al\be} [{\bar c}\ga_\mu P_R q][{\bl}_\al\ga^\mu P_L\nu_\be] \right. ~ \nl
&&\hspace{1truecm} +~(\epsilon_S)_{\al\be}[{\bar c} q][{\bl}_\al P_L\nu_\be] + (\epsilon_P)_{\al\be}[{\bar c}\ga^5 q][{\bl}_\al P_L\nu_\be] ~\nl
&&\hspace{5truecm} \left.+~(\epsilon_T)_{\al\be}[{\bar c}\si_{\mu\nu} P_L q][\bl_\al\si^{\mu\nu} P_L\nu_\be]\right\} \hc,~~
\eea
where $\al,\be$ are lepton flavor indices representing the three flavors $e, \mu$, and $\tau$, and $q = s, d$. 
Here $P_R$ and $P_L$ are respectively the right and left projection operators $(1 \pm \ga^5)/2$, $G_F$ refers to the Fermi constant, $V_{cq}$ refers to the appropriate CKM matrix element, and $(\epsilon_X)_{\al\be}$ refer to the NP Wilson coefficients where $X$ can be $L, R, S, P$, or $T$ for left-handed vector, right-handed vector, scalar, pseudoscalar, or tensor. Here the neutrinos are left-handed Dirac particles and $X$ refers to the Lorentz structure of the quark current.

To obtain benchmarks for $(\epsilon_X)_{\al\be}$ we will study processes where the quark-level transition is of the type $({\bar c}q)(\bl\nu)$, specifically two- and three-body decays of $D$ and $D^*$ mesons. Now, since neutrinos are not directly detected at collider experiments the final state neutrino flavor in a two- or three-body meson decay is unknown. Any effect of neutrino flavor gets washed out in summing over the final-state neutrino flavor. While in principle NSI can lead to lepton flavor violation when $(\epsilon_X)_{\al\be}$ is nonzero for $\al\ne\be$, here we will always sum over the neutrino flavor index $\be$. The resulting Hamiltonian is effectively flavor diagonal and can be expressed as
\bea
\label{eq:Heffn}
H_{\rm eff} &=& \frac{4G_FV_{cq}}{\sqrt{2}}\left\{(1+\epsilon_L)[{\bar c}\ga_\mu P_L q][{\bl}\ga^\mu P_L\nu] + \epsilon_R [{\bar c}\ga_\mu P_R q][{\bl}\ga^\mu P_L\nu] \right. ~ \nl
&&\hspace{2truemm} \left. +~\epsilon_S [{\bar c} q][{\bl}_\al P_L\nu_\be] + \epsilon_P[{\bar c}\ga^5 q][\bl P_L\nu] + \epsilon_T[{\bar c}\si_{\mu\nu} P_L q][\bl\si^{\mu\nu} P_L\nu]\right\} \hc ,~~
\eea
where without the loss of generality here and in what follows we have suppressed the charge-lepton flavor index $\alpha$ on the effective WCs $\epsilon_X$ ($X = L,R,S,P,T$) obtained after summing over the neutrino flavor index $\be$,
\beq
\epsilon_X^\al ~=~ \sum\limits_\be^{e,\mu,\tau}(\epsilon_X)_{\al\be}~.~~
\eeq

It is common to use a different nomenclature and basis of WCs, especially for the scalar and pseudoscalar operators which can instead be written in terms of left- and right-handed operators. This is done by rewriting the above Hamiltonian as follows
\bea
\label{eq:Heff}
H_{\rm eff} &=& \frac{4G_FV_{cq}}{\sqrt{2}}\left\{(1+V_L)[{\bar c}\ga_\mu P_L q][{\bl}\ga^\mu P_L\nu] + V_R [{\bar c}\ga_\mu P_R q][{\bl}\ga^\mu P_L\nu] \right. ~ \nl
&&\hspace{2truemm} \left. +~S_L [{\bar c}P_L q][{\bl} P_L\nu] + S_R[{\bar c}P_R q][\bl P_L\nu] + T_L[{\bar c}\si_{\mu\nu} P_L q][\bl\si^{\mu\nu} P_L\nu]\right\} \hc ,~~
\eea
where $X_Y$ represent the effective NP WCs with $X = S, V, T$ referring to scalar, vector, or tensor and $Y = L, R$ referring to left-handed and right-handed quark currents. We have once again suppressed the charged-lepton flavor index, $\al$, in Eq.~(\ref{eq:Heff}). By comparing Eqs.~(\ref{eq:Heffn}) and (\ref{eq:Heff}) one can relate the two sets of WCs as follows,
\beq
V_{L(R)} ~=~ \epsilon_{L(R)} ~,~~ S_L ~=~ \epsilon_S - \epsilon_P ~,~~ S_R ~=~ \epsilon_S + \epsilon_P~.~~T_L ~=~ \epsilon_T ~,~~
\eeq
Thus, constraints from meson decay experiments can be directly translated into bounds on five WCs, $V_L, V_R, S_L, S_R$, and $T_L$ for each lepton flavor $e, \mu$, and $\tau$.

The effective Hamiltonian given in Eq.~\eqref{eq:Heff} above can come from various ultraviolet-complete models. We discuss two specific examples -- the leptoquark model and the vector boson ($W'$) model. Here we demonstrate how the Wilson coefficients in Eq.~\eqref{eq:Heff} can be expressed in terms of tree-level couplings within these ultraviolet-complete models.

The interaction of a singlet leptoquark, $S_1(\bar{3},1,1/3)$, with SM fermions can be expressed as \cite{Buchmuller:1986zs},
\beq
\cL^{\rm LQ} = \left(g_{1L}^{ij}\,\bar{Q}_{iL}^c i\sigma_2 L_{jL} + g_{1R}^{ij}\,\bar{u}_{iR}^c \ell_{jR} \right)S_1 \hc, \
\eeq
where $Q_i$ and $L_j$ represent the left-handed quark and lepton doublets, $u_{iR}$ and $d_{iR}$ are the right-handed up-type and down-type quark singlets and $\ell_{jR}$ represents the right-handed charged lepton singlets. Indices $i$ and $j$ denote the generations of quarks and leptons. Integrating out the heavy $S_1$ leptoquark one obtains the following effective operators involving neutrinos.
\bea
\cL^{(1)}_\eff &=& \frac{g^{ij}_{1L}g^{kl*}_{1L}}{M^2_{S_1}}\(\bar{Q}_{iL}^c \si_2 L_{jL}\)
\({\bL}_{kL}\si_2Q^c_{lL}\) ,~~ \\
\cL^{(2)}_\eff &=& \frac{g^{ij}_{1L}g^{kl*}_{1R}}{M^2_{S_1}}\(\bar{Q}_{iL}^c i\sigma_2 L_{jL}\)\({\bl}_{lR}u^c_{kR}\) \hc ,~~ \\
&=& \frac{g^{ij}_{1L}g^{kl*}_{1R}}{M^2_{S_1}}\lb\({\bar u}^c_{iL}\ell^{}_{jL}\)\({\bl}_{lR}u^c_{kR}\) - \({\bar d}^c_{iL}\nu^{}_{jL}\)\({\bl}_{lR}u^c_{kR}\)\rb \hc ~~
\eea
After a Fierz transformation, $\cL^{(1)}_\eff$ can be written as
\bea
\cL^{(1)}_\eff &=& \frac{g^{ij}_{1L}g^{kl*}_{1L}}{4M^2_{S_1}}\[\(\bar{Q}_{kL}\ga^\mu Q_{iL}\)\(\bL_{lL}\ga_\mu L_{jL}\) - \(\bar{Q}_{kL}\ga^\mu\si^I Q_{iL}\)\(\bL_{lL}\ga_\mu\si^I L_{jL}\)\] .~~
\eea
Only the second term above contains lepton-neutrino interactions which can be expressed using the following effective Hamiltonian.
\bea
\cH^{(1)}_\eff &=&-~\frac{g^{2l}_{1L}g^{2l*}_{1L}}{2M^2_{S_1}}\({\bar c} \ga^\mu P_L s\)\(\bl\ga_\mu P_L\nu\) \hc
\eea

In a similar vein, the second of the two terms in $\cL^{(2)}_\eff$, after Fierz transformation, gives rise to the following effective Hamiltonian,
\bea
\cH^{(2)}_\eff &=& -\frac{g^{2l}_{1L} g^{2l*}_{1R}}{2M^2_{S_1}}\[\({\bar c}P_Ls\)\({\bl}P_L\nu\) + 
\frac{1}{4}\({\bar c}\si^{\mu\nu}P_Ls\)\({\bl}\si^{}_{\mu\nu}P_L\nu\)\] \hc ~~
\eea
 Comparing the above effective Hamiltonians with that of Eq.~(\ref{eq:Heff}) and restricting ourselves to second-generation quarks, we can express the coefficients $S_L$, $V_L$, and $T_L$ as follows,
\beq
S_L = -\frac{1}{2\sqrt2 G_F V_{cs}} \frac{g_{1L}^{2l}g_{1R}^{2l*}}{2M_{S_1}^2}, ~~ V_L = \frac{1}{2\sqrt2 G_F V_{cs}}\frac{g_{1L}^{2l}g_{1L}^{2l*}}{2M_{S_1}^2} ,~~ T_L = -\frac{1}{2\sqrt2 G_F V_{cs}} \frac{g_{1L}^{2l}g_{1R}^{2l*}}{8M_{S_1}^2}. \,
\eeq

Let us now consider the following Lagrangian, encoding the interaction of a $W'$ boson with leptons and quarks:
\bea
\cL^{W'} &=& \frac{g}{\sqrt{2}}\[V_{ij} \lb\bar{u}_i \ga^\mu( g^{ij}_L P_L +  g^{ij}_R P_R) d_j\rb + g^{\nu_\ell\ell}\(\bl\ga^\mu P_L\nu_\ell\)\]W^{\prime+}_\mu \hc ,~~
\eea
where $g$ is the SM Weak coupling, $V_{ij}$ represents the relevant CKM matrix element, $g^{ij}_X$ ($X = L,R$) and $g^{\nu_\ell\ell}$ are the NP couplings of the $W'$ boson with the SM quarks and leptons. Using the definition of the fermi constant, $G_F/\sqrt{2} = g^2/8m^2_W$, and integrating out the $W'$ leads to the following effective Hamiltonian.
\bea
\cH_\eff &=&  \frac{4 G_FV_{cq} }{\sqrt{2}}
\[\bar{c}_i\ga^\mu\(  \frac{M_W^2}{M_{W'}^2}g^{cq}_L P_L + \frac{M_W^2}{M_{W'}^2} g^{cq}_R P_R\) q\]
\[g^{\nu_\ell\ell}\bl \ga_\mu P_L\nu_\ell\] \hc ~~
\label{intw}
\eea
Comparing Eq.~\eqref{intw} with Eq.~\eqref{eq:Heff}  we obtain the following relations:
\beq
V_L ~=~ \frac{M_W^2}{M_{W'}^2}g^{cq}_L\sum\limits_{\nu_\ell} g^{\nu_\ell\ell},~~~~~
V_R ~=~ \frac{M_W^2}{M_{W'}^2}g^{cq}_R \sum\limits_{\nu_\ell}g^{\nu_\ell\ell}.
\label{LR}
\eeq

\section{Flavor Observables}
In this section, we discuss the low-energy fits to the WCs of the effective Lagrangian from low-energy data and select a few benchmarks for NP. There are two main types of observables, listed below, that we are interested in for benchmarking.
\bit
 \item \textbf{Two-body decays:} $D_s^+\to\ell^+\nu,~ D^+\to\ell^+\nu$
 \item \textbf{Three-body decays:} $D\to K\ell\nu$,~ $D\to\pi\ell\nu$, ~$D\to K^*\ell\nu$,~ $D\to\rho\ell^+\nu$, ~$D_s^+\to\phi\ell^+\nu$, ~$D_s^+\to\eta^{(\prime)\ell\nu}$,~$\Lambda_c \to \Lambda \ell \nu$
\eit
In the following sections, we will discuss benchmarks obtained from the above list.
In our analysis we will not include the weak decays of the vector mesons as they decay dominantly by the strong and electromagnetic interactions. Further we will ignore the tensor current contributions to the pseudoscalar charmed mesons and charmed baryon decays as the relevant form factors are not known very well.

\subsection{Two-body meson decays}
The decay rate for the process $M^+_{ij}\to\ell^+\nu$, where $M^+_{ij}$ represents a pseudoscalar $q^i_u\bqd^j$ meson, can be expressed in the basis of Eq.~\eqref{eq:Heff} as \cite{Becirevic:2020rzi} :
\bea
&\Ga_{M^+_{ij}\to\ell^+\nu} ~=~ \frac{G^2_F m_M m^2_\ell}{8\pi}|V_{ij}|^2f^2_M\(1-\frac{m^2_\ell}{m^2_M}\)^2 \lb |1 + V_L - V_R|^2 + \frac{|S_L - S_R|^2m^4_M}{m^2_\ell(m_i + m_j)^2} \rb ,~~
\eea
where $m_i$ is the mass of the $q^i_{u,d}$, $V_{ij}$ represents the relevant CKM matrix element, and other symbols carry their usual meaning. All numerical inputs are listed in Table~\ref{tab:input} in Appendix~\ref{app:formfactors}. The SM predictions using these parameters are given in Table~\ref{tab:obs}.

\subsection{Three-body meson decays}
The decay rate of the semileptonic three body meson decays $M_i \to M_j \ell \nu$ in the basis of the Hamiltonian in Eq.~\eqref{eq:Heff} can be written as \cite{Sakaki:2013bfa, Bhattacharya:2022bdk}:
\bea
\frac{d\Gamma}{dq^2} &=& \frac{G_F^2 |V_{ij}|^2}{192 \pi^3 m_{M_i}^3} q^2 \sqrt{\lambda(m_{M_i}^2, m_{M_j}^2, q^2)}\(1-\frac{m_\ell^2}{q^2}\)^2\times \nl
&& \bigg \{ |1+V_L + V_R|^2 \[\(1+ \frac{m_\ell^2}{2q^2}\) h_{V,0}^{2}(q^2) +~\frac{3}{2} \frac{m_\ell^2}{q^2}h_{V,t}^{~2}(q^2)\] +\frac{3}{2}|S_L + S_R|^2 h_S^2(q^2) \nl
&&  +~3 \mathcal{R}\rm{e}\[(1+V_L+V_R)(S_L^* + S_R^*)\] \frac{m_\ell}{\sqrt{q^2}}h_S(q^2) h_{V,t}(q^2) \bigg \},
\eea
where $\lambda(x,y,z) = x^2 + y^2 + z^2 - 2(xy+yz+zx)$.

The semileptonic three-body decay rate for $M_i \to M_j^* \ell \nu$ decays is expressed as:
\bea
\frac{d\Gamma}{dq^2} &=& \frac{G_F^2 |V_{ij}|^2}{192\pi^3 m_{M_i}^3}q^2 \sqrt{\lambda(m_{M_i}^2, m_{M_j^*}^{2}, q^2)}\(1-\frac{m_\ell^2}{q^2}\)^2 \times \nl
&& \bigg \{(|1+V_L|^2 + |V_R|^2)\(\(1+\frac{m_\ell^2}{2q^2}\)(H_{V,+}^2(q^2) + H_{V,-}^2(q^2) +H_{V,0}^2(q^2)) + \frac{3}{2}\frac{m_\ell^2}{q^2}H_{V,t}^2\) \nl
&& -2\mathcal{R}e\[(1+V_L)V_R^*\]\(\(1+\frac{m_\ell^2}{2q^2}\)(2H_{V,+}^2(q^2) H_{V,-}^2(q^2) + H_{V,0}^2(q^2)) + \frac{3}{2}\frac{m_\ell^2}{q^2}H_{V,t}^2(q^2)\)\nl
&& +\frac{3}{2}|S_L-S_R|^2 H_S^2(q^2) + 3\mathcal{R}e\[(1+V_L-V_R)(S_L^*-S_R^*)\]\frac{m_\ell}{\sqrt{q^2}}H_S(q^2) H_{V,t}(q^2)\bigg \}.
\eea 

The above hadronic amplitudes in terms of the form factors are provided in Appendix~\ref{app:hadamp} and the form factor parameters and other numerical inputs are detailed in Appendix~\ref{app:formfactors}.

\subsection{Three-body baryonic decays}

In this work we also consider the semileptonic decay of charmed baryon $\Lambda_c \to \Lambda \ell \nu$. The decay rate in terms of new physics Wilson coefficients as defined in Eq.~\eqref{eq:Heff} has been adopted from  Ref.~\cite{Shivashankara:2015cta,Datta:2017aue} for the $\Lambda_b \to \Lambda_c \ell \nu$ decay as :
\beq 
\frac{d\Gamma}{dq^2} = \frac{G_F^2|V_{cs}|^2q^2}{384\pi^3 m_{\Lambda_c}^3}\sqrt{Q_+ Q_-}\(1-\frac{m_\ell^2}{q^2}\)^2\(A_1^{VA}+\frac{m_\ell^2}{2q^2}A_2^{VA}+\frac{3}{2}A_3^{SP}+\frac{3m_\ell}{\sqrt{q^2}}A_5^{VA-SP}\),
\eeq 
where $Q_{\pm} = (m_{\Lambda_c} \pm m_\Lambda)^2-q^2$ and the amplitudes $A_i^X$ in terms of the form factors and WCs can be found in Ref.~\cite{Datta:2017aue}.

\begin{table}[t]
\centering
\footnotesize
\resizebox{\textwidth}{!}{
\begin{tabular}{|c|c|c|c|c|c|c|}\hline
\multirow{2}{*}{Decay}  & \multicolumn{2}{c|}{$\ell = e$} & \multicolumn{2}{c|}{$\ell = \mu$}& \multicolumn{2}{c|}{$\ell = \tau$}\\
\cline{2-7}
&$\mathcal{B}^{\rm{SM}}$  (\%) & $\mathcal{B}^{\rm{meas}}$ (\%) & $\mathcal{B}^{\rm{SM}}$  (\%) & $\mathcal{B}^{\rm{meas}}$ (\%) & $\mathcal{B}^{\rm{SM}}$  (\%) & $\mathcal{B}^{\rm{meas}}$ (\%)\\
\hline \hline
$D^+_s\to\ell^+\nu_\ell$ & $(1.28 \pm 0.05)\times 10^{-5}$ & $< 8.3\times 10^{-3}$ & $0.546 \pm 0.020$ & $0.543\pm 0.015$  & $5.32 \pm 0.20$ & $5.32 \pm 0.11$ \\
$D^+\to\ell^+\nu_\ell$ & $(8.85 \pm 0.47) \times 10^{-7}$ & $< 8.8\times 10^{-4}$ & $0.0376 \pm 0.0020$ & $0.0374\pm 0.0017$  & $0.100 \pm 0.005$ & $0.120 \pm 0.027$ \\
$D^+\to \overline{K}^0 \ell^+\nu_\ell$ & $8.99 \pm 0.37$ & $8.72\pm 0.09$ & $8.77 \pm 0.36$ & $8.76\pm 0.19$ &&\\
$D^0\to K^- \ell^+ \nu_\ell$ & $3.54 \pm 0.15$ & $3.549 \pm 0.026$ & $3.45 \pm 0.14$ & $3.41\pm 0.04$ &&\\
$D^+\to \pi^0 \ell^+\nu_\ell$ & $0.320 \pm 0.031$& $0.372 \pm 0.017$ & $0.316 \pm 0.030$ & $0.350\pm 0.015$  &&\\
$D^0\to \pi^- \ell^+ \nu_\ell$ & $0.249 \pm 0.024$ & $0.291 \pm 0.004$ & $0.245 \pm 0.023$ & $0.267\pm 0.012$  && \\
$D^+\to \overline{K}^{*}(892)^0 \ell^+\nu_\ell$ & $5.84 \pm 1.61$ & $5.40\pm 0.10$ & $5.52 \pm 1.52$ & $5.27\pm 0.15$  && \\
$D^0\to K^{*}(892)^- \ell^+\nu_\ell$ & $2.30 \pm 0.64$ & $2.15\pm 0.16$ & $2.18 \pm  0.60$ & $1.89\pm 0.24$ && \\
$D^+\to \rho^0 \ell^+\nu_\ell$ & $0.212 \pm 0.048$ & $0.19 \pm 0.01$&  $0.203 \pm 0.046$ & $0.24\pm 0.04$ &&\\
$D^0\to \rho^- \ell^+\nu_\ell$ & $0.171 \pm 0.039$ & $0.150\pm 0.012$ & $0.163 \pm 0.037$& $0.135\pm 0.013$  &&\\
$D^+_s\to\phi\ell^+\nu_\ell$ & $2.68 \pm 0.26$ & $2.39\pm 0.16$ & $2.53 \pm 0.23$ & $1.9\pm 0.5$ &&\\
$D^+_s\to\eta\ell^+\nu_\ell$ & $2.96 \pm 0.45$ & $2.26\pm 0.05$ & $2.91 \pm 0.45$ & $2.4\pm 0.5$ && \\
$D^+_s\to\eta^\prime(958)\ell^+\nu_\ell$ & $0.909 \pm 0.41$ & $0.8\pm 0.04$ & $0.869 \pm 0.41$  & $1.1\pm0.5$ &&\\
$\Lambda_c^+ \to \Lambda \ell^+ \nu_\ell$ & $3.92 \pm 0.63$ & $3.56 \pm 0.13$ & $3.80 \pm 0.60$ & $3.48 \pm 0.17$ &&\\
\hline \hline
\end{tabular}}
\caption{List of observables used to constrain the new physics coefficients along with their SM predictions and measured branching fractions. The measured values are taken from the PDG \cite{ParticleDataGroup:2022pth} while the SM predictions are computed using the form factors defined in Appendix~\ref{app:formfactors} and the CKM matrix elements listed in Table~\ref{tab:input}.}
\label{tab:obs}
\end{table}

\subsection{Low energy fit results}
\label{sec:lowfitresults}

We fit the new physics Wilson coefficients in Eq.~\eqref{eq:Heff} to the observables listed in Table~\ref{tab:obs} separately for the electron and muon modes. In Table~\ref{tab:obs}, we list the SM expectation and the measured values of the branching fractions for each decay mode. For our numerical analysis, we take the values of $D$-meson decay constants and lattice determinations of relevant CKM matrix elements from the FLAG Review \cite{FlavourLatticeAveragingGroupFLAG:2021npn} \footnote{Ref.~\cite{Bolognani:2024cmr} points out that the inclusion of a universal electroweak correction in the Wilson coefficient $\epsilon_L$ leads to a $\sim$ 1\% increase in $\epsilon_L$. This also leads to a smaller value of $|V_{cs}|$ compared to the one obtained by the PDG \cite{ParticleDataGroup:2022pth}. In this work, however, we use $\epsilon_L = 1$ and $|V_{cs}|$ as given in the FLAG Review \cite{FlavourLatticeAveragingGroupFLAG:2021npn}.}. In Appendix~\ref{app:formfactors} we list the relevant numerical inputs and form factors used in this article. Note that there are 12 measurements for the electron modes and 14 for the muon modes which are utilized to fit to the respective flavour specific Wilson coefficients. For the tau mode, however, we only have measurements for the two body leptonic decay of the $D_{(s)}$ meson. Furthermore, since the $\nu_\tau$ detection efficiency at SND is low compared to the other lepton flavour, it is challenging to obtain any meaningful constraint on the new physics parameters. Hence, we do not fit to tau measurements in our analysis. 

The allowed parameter values
for a specific model X corresponding to a specific lepton flavour is determined using
\beq
\chi^2(X) = \sum_{i=1}^{n}\frac{(\mathcal{B}_i^{\rm{th}}(X)-\mathcal{B}_i^{\rm{meas}})^2}{\sigma_i^2},
\eeq 
where n is the total number of observables, $\mathcal{B}^{\rm{th}}$ is the theoretical branching fraction, $\mathcal{B}^{\rm{meas}}$ is the corresponding experimental measurement and $\sigma$ is the uncertainty from measurement and theory added in quadrature.
For a given model, the set of Wilson coefficients that minimize the above $\chi^2$ function are found 
with their respective $1\sigma$ uncertainties using the algorithm supplied by the MINUIT package~\cite{James:1975dr,iminuit} 
as shown in Tables~\ref{tab:electronfit},~\ref{tab:muonfit}. We fit to both real and imaginary components of the Wilson coefficients denoted as $\epsilon_X = \Re[\epsilon_{X}] + i\,\Im[\epsilon_{X}]$ where $X$ can be $L,R,S,P$. In each case, we define the pull with respect to the SM as $\sqrt{\chi^2(SM)-\chi^2(X)}$.

\begin{table}[t!!]
\centering
\renewcommand{\arraystretch}{1.4}
\resizebox{\textwidth}{!}{\begin{tabular}{|c|c|c|c|c|c|c|c|c|c|c|c|}
\hline
Model & Fit Parameters & $\chi^2$/dof & $\Re\[\epsilon_{L}\]$ & $\Im\[\epsilon_{L}\]$ & $\Re\[\epsilon_{R}\]$ & $\Im\[\epsilon_{R}\]$ & $\Re\[\epsilon_{S}\]$ & $\Im\[\epsilon_{S}\]$ & $\Re\[\epsilon_{P}\]$ & $\Im\[\epsilon_{P}\]$ & pull \\
\hline
\multirow{2}*{1} & $\Re\[\epsilon_{L}\], \Im\[\epsilon_{L}\], \Re\[\epsilon_{R}\], \Im\[\epsilon_{R}\],$   & \multirow{2}*{1.18/4} & \multirow{2}*{$-1.14(13)$} & \multirow{2}*{$-0.54(14)$} & \multirow{2}*{$-0.56(13)$} & \multirow{2}*{$0.32(14)$} & \multirow{2}*{$-0.53(14)$} & \multirow{2}*{$-0.17(40)$} & \multirow{2}*{$0.0(1.2)$} & \multirow{2}*{$0.0(1.2)$} & \multirow{2}*{2.95} \\
& $\Re\[\epsilon_{S}\], \Im\[\epsilon_{S}\], \Re\[\epsilon_{P}\], \Im\[\epsilon_{P}\]$ & & & & & & & & & &\\
\hline
2 & $\Re\[\epsilon_{L}\], \Re\[\epsilon_{R}\], \Re\[\epsilon_{S}\], \Re\[\epsilon_{P}\]$ &1.36/8 & $-0.877(60)$ & -- & $-0.83(7)$ & -- & $0.58(10)$ & -- & $0.0(1.2)$ & -- & 2.92\\
\hline
3 & $\Re\[\epsilon_{L}\], \Im\[\epsilon_{L}\]$ & 9.8/10 & $-0.0043(35)$ & $0.014(25)$ & -- & -- & -- & -- & -- & -- & 0.32 \\
\hline
4 & $\Re[\epsilon_{L}]$ & 9.8/11 & $-0.0042(130)$ & -- & -- & -- & -- & -- & -- & -- & 0.32 \\
\hline
5 & $\Re[\epsilon_{R}], \Im[\epsilon_{R}]$ & 9.7/10 & -- & -- & $0.006(13)$ & $0.00(13)$ & -- & -- & -- & -- & 0.45 \\
\hline
6 & $\Re[\epsilon_{R}]$ & 9.7/11 & -- & -- & $0.006(13)$ & -- & -- & -- & -- & -- & 0.45 \\
\hline
7 & $\Re[\epsilon_{L}], \Im[\epsilon_{L}], \Re[\epsilon_{R}], \Im[\epsilon_{R}]$ & 8.2/8 & $-0.32(12)$ & $-0.70(19)$ & $0.021(90)$ & $-0.21(26)$ & -- & -- & -- & -- & 1.3\\
\hline
8 & $\Re[\epsilon_{L}], \Re[\epsilon_{R}]$ & 8.2/10 & $-0.028(24)$ & -- & $0.03(2)$ & -- & -- & -- & -- & -- & 1.3\\
\hline
9 & $\Re[\epsilon_{S}], \Im[\epsilon_{S}]$ & 8.15/5 & -- & -- & -- & -- & $-0.098(230)$ & $0.0(9)$ & -- & -- & 0.52\\ 
\hline
10 & $\Re[\epsilon_{S}]$ & 8.15/6 & -- & -- & -- & -- & $-0.098(230)$ & -- & -- & -- & 0.52\\ 
\hline
11 & $\Re[\epsilon_{P}], \Im[\epsilon_{P}]$ &1.79/4 & -- & -- & -- & -- & -- & -- & $0.0(6)$ & $0.0(6)$ & 0.004\\
\hline
12 & $\Re[\epsilon_{P}]$ &1.79/5 & -- & -- & -- & -- & -- & -- & $0.0(6)$ & -- & 0.004\\
\hline
13 & $\Re[\epsilon_{S}], \Im[\epsilon_{S}], \Re[\epsilon_{P}], \Im[\epsilon_{P}]$ & 9.63/8 & -- & -- & -- & -- & $-0.098(290)$ & $0.0(8)$ & $-0.002(600)$ & $0.0(6)$ & 0.52 \\
\hline
14 & $\Re[\epsilon_{S}], \Re[\epsilon_{P}]$ & 9.63/10 & -- & -- & -- & -- & $-0.098(290)$ & -- & $-0.002(600)$ & -- & 0.52 \\
\hline
\multirow{2}*{15} & $\Re[\epsilon_{L}], \Im[\epsilon_{L}], \Re[\epsilon_{R}], \Im[\epsilon_{R}],$ & \multirow{2}*{1.18/6} & \multirow{2}*{$-0.73(60)$} & \multirow{2}*{$-0.49(17)$} & \multirow{2}*{$0.46(60)$} & \multirow{2}*{$0.45(18)$} & \multirow{2}*{$0.56(34)$} & \multirow{2}*{$-0.032(180)$} & \multirow{2}*{--} & \multirow{2}*{--} & \multirow{2}*{2.95}\\
 & $\Re[\epsilon_{S}], \Im[\epsilon_{S}]$ & & & & & & & & & & \\
\hline
16 & $\Re[\epsilon_{L}], \Re[\epsilon_{R}], \Re[\epsilon_{S}]$ & 1.35/8 & $-0.88(6)$ & -- & $-0.83(7)$ & -- & $-0.58(11)$ & -- & -- & -- & 2.92\\
\hline
\multirow{2}*{17} & $\Re[\epsilon_{L}], \Im[\epsilon_{L}], \Re[\epsilon_{R}], \Im[\epsilon_{R}],$ & \multirow{2}*{8.2/6} & \multirow{2}*{$-0.54(14)$} & \multirow{2}*{$0.86(18)$} & \multirow{2}*{$0.014(130)$} & \multirow{2}*{$0.026(180)$} & -- & -- & \multirow{2}*{$0.0(4)$} & \multirow{2}*{$0.0(3)$} & \multirow{2}*{1.3} \\
& $\Re[\epsilon_{P}], \Im[\epsilon_{P}]$ & & & & & & & & & & \\
\hline
18 & $\Re[\epsilon_{L}], \Re[\epsilon_{R}], \Re[\epsilon_{P}]$ & 8.2/9 & $-0.028(23)$ & -- & $0.030(24)$ & -- & -- & -- & $0.0(4)$ & -- & 1.3\\
\hline
\end{tabular}}
\caption{Fit results for different combinations of WCs in decays to electrons. The SM chi-square for the full set of observables listed in Table~\ref{tab:obs} is $\chi^2_{SM}/dof = 9.90/12$ while for those affecting the scalar coupling only, $\chi^2_{SM,S}/dof = 8.43/7$, and correspondingly for the pseudoscalar coupling, $\chi^2_{SM,P}/dof = 1.79/6$. }
\label{tab:electronfit}
\end{table}

Note that, while all observables considered in this work are sensitive to vector-type new physics, not all decay rates are affected by the scalar coupling $\epsilon_S$ or pseudoscalar coupling $\epsilon_P$. The two body leptonic decay of pseudoscalar mesons $M_i \to \ell \nu$ are purely sensitive to $\epsilon_P$ while that for the vector mesons are sensitive to neither $\epsilon_S$ nor $\epsilon_P$. The three body semileptonic decays of the form $M_i \to M_j \ell \nu$ are affected by $\epsilon_S$ only while $M_i \to M_j^* \ell \nu$ rates are sensitive to $\epsilon_P$ only. The charmed baryonic decay $\Lambda_c \to \Lambda \ell \nu$, however, are influenced by both scalar and pseudoscalar new physics. Hence, while performing the fits to models involving either $\epsilon_S$ or $\epsilon_P$ only, we drop the decay modes that are redundant to the analysis for that particular coupling. The SM chi-squares are calculated accordingly in order to estimate the pull with respect to new physics. As mentioned in the caption of Table~\ref{tab:electronfit}, the SM chi-square value per degree of freedom ($dof$) for the full set of 12 observables listed in Table~\ref{tab:obs} is calculated to be $\chi^2_{SM}/dof = 9.9/12$ while, for the observables sensitive to $\epsilon_S$ only, $\chi^2_{SM,S}/dof = 8.43/7$, and, for those affected by $\epsilon_P$, $\chi^2_{SM,P}/dof = 1.79/6$. Similarly, for the muon fit, the respective SM chi-squares are found to be $\chi^2_{SM}/dof = 4.94/14$, $\chi^2_{SM,S}/dof = 2.5/7$ and $\chi^2_{SM,P}/dof = 2.7/8$.

The pseudoscalar coupling $\epsilon_P$ is tightly constrained from the two body decays $D_s^+ \to \ell^+ \nu_\ell$ and $D^+ \to \ell^+ \nu_\ell$. However, the lack of measurement of these decay rates for the electron channel keeps $\epsilon_P$ practically unconstrained as is reflected in the fit results in Table~\ref{tab:electronfit}. In case of the muon, $\epsilon_P$ is better constrained due to the availability of the measurements of the two body decays and we obtained tightly constrained central values with small error bars as shown in Table~\ref{tab:muonfit}. For the scalar and vector couplings, there is an interplay between the semileptonic mesonic three body decays and the charmed baryon decays resulting in very small to even $\mathcal{O}(1)$ central values for the real and imaginary parts of the parameters.

\begin{table}[t!!]
\centering
\renewcommand{\arraystretch}{1.4}
\resizebox{\textwidth}{!}{
\begin{tabular}{|c|c|c|c|c|c|c|c|c|c|c|c|}
\hline
Model & Fit Parameters & $\chi^2$/dof & $\Re\[\epsilon_{L}\]$ & $\Im\[\epsilon_{L}\]$ & $\Re\[\epsilon_{R}\]$ & $\Im\[\epsilon_{R}\]$ & $\Re\[\epsilon_{S}\]$ & $\Im\[\epsilon_{S}\]$ & $\Re\[\epsilon_{P}\]$ & $\Im\[\epsilon_{P}\]$ & pull \\
\hline
\multirow{2}*{1} & $\Re\[\epsilon_{L}\], \Im\[\epsilon_{L}\], \Re\[\epsilon_{R}\], \Im\[\epsilon_{R}\],$   & \multirow{2}*{2.08/6} & \multirow{2}*{$-1.71(10)$} & \multirow{2}*{$0.36(14)$} & \multirow{2}*{$-0.14(9)$} & \multirow{2}*{$-0.39(14)$} & \multirow{2}*{$-0.36(13)$} & \multirow{2}*{$-0.011(8)$} & \multirow{2}*{$0.007(8)$} & \multirow{2}*{$-0.009(7)$} & \multirow{2}*{1.69} \\
& $\Re\[\epsilon_{S}\], \Im\[\epsilon_{S}\], \Re\[\epsilon_{P}\], \Im\[\epsilon_{P}\]$ & & & & & & & & & &\\
\hline
2 & $\Re\[\epsilon_{L}\], \Re\[\epsilon_{R}\], \Re\[\epsilon_{S}\], \Re\[\epsilon_{P}\]$ &2.14/10 & $-0.097(10)$ & -- & $-0.049(14)$ & -- & $0.35(16)$ & -- & $-0.011(9)$ & -- & 1.67\\
\hline
3 & $\Re\[\epsilon_{L}\], \Im\[\epsilon_{L}\]$ & 4.87/12 & $-1.98(1.10)$ & $-0.18(1.70)$ & -- & -- & -- & -- & -- & -- & 0.26 \\
\hline
4 & $\Re[\epsilon_{L}]$ & 4.87/13 & $-2.00(18)$ & -- & -- & -- & -- & -- & -- & -- & 0.26 \\
\hline
5 & $\Re[\epsilon_{R}], \Im[\epsilon_{R}]$ & 4.83/12 & -- & -- & $0.0038(110)$ & $0.0(2)$ & -- & -- & -- & -- & 0.34 \\
\hline
6 & $\Re[\epsilon_{R}]$ & 4.83/12 & -- & -- & $0.0038(110)$ & -- & -- & -- & -- & -- & 0.34 \\
\hline
7 & $\Re[\epsilon_{L}], \Im[\epsilon_{L}], \Re[\epsilon_{R}], \Im[\epsilon_{R}]$ & 4.70/10 & $-1.91(28)$ & $-0.41(15)$ & $0.00(28)$ & $0.00(15)$ & -- & -- & -- & -- & 0.48\\
\hline
8 & $\Re[\epsilon_{L}], \Re[\epsilon_{R}]$ & 4.70/12 & $-0.004(12)$ & -- & $0.005(12)$ & -- & -- & -- & -- & -- & 0.48\\
\hline
9 & $\Re[\epsilon_{S}], \Im[\epsilon_{S}]$ & 1.24/5 & -- & -- & -- & -- & $-0.31(6)$ & $0.00(34)$ & -- & -- & 1.12\\ 
\hline
10 & $\Re[\epsilon_{S}]$ & 1.24/6 & -- & -- & -- & -- & $-0.31(6)$ & -- & -- & -- & 1.12\\ 
\hline
11 & $\Re[\epsilon_{P}], \Im[\epsilon_{P}]$ &2.70/6 & -- & -- & -- & -- & -- & -- & $-0.00014(13)$ & $0.00000(13)$ & 0.01\\
\hline
12 & $\Re[\epsilon_{P}]$ &2.70/7 & -- & -- & -- & -- & -- & -- & $-0.00014(13)$ & -- & 0.01\\
\hline
13 & $\Re[\epsilon_{S}], \Im[\epsilon_{S}], \Re[\epsilon_{P}], \Im[\epsilon_{P}]$ & 3.68/10 & -- & -- & -- & -- & $-0.31(16)$ & $0.00(33)$ & $-0.00013(13)$ & $0.00000(13)$ & 1.12 \\
\hline
14 & $\Re[\epsilon_{S}], \Re[\epsilon_{P}]$ & 3.68/12 & -- & -- & -- & -- & $-0.31(16)$ & -- & $-0.00013(13)$ & -- & 1.12 \\
\hline
\multirow{2}*{15} & $\Re[\epsilon_{L}], \Im[\epsilon_{L}], \Re[\epsilon_{R}], \Im[\epsilon_{R}],$ & \multirow{2}*{2.65/8} & \multirow{2}*{$-0.11(26)$} & \multirow{2}*{$0.09(15)$} & \multirow{2}*{$-0.057(80)$} & \multirow{2}*{$-0.19(27)$} & \multirow{2}*{$0.36(31)$} & \multirow{2}*{$-0.04(40)$} & \multirow{2}*{--} & \multirow{2}*{--} & \multirow{2}*{1.51}\\
& $\Re[\epsilon_{S}], \Im[\epsilon_{S}]$ & & & & & & & & & & \\
\hline
16 & $\Re[\epsilon_{L}], \Re[\epsilon_{R}], \Re[\epsilon_{S}]$ & 2.65/10 & $-0.082(40)$ & -- & $-0.075(40)$ & -- & $0.36(15)$ & -- & -- & -- & 1.51\\
\hline
\multirow{2}*{17} & $\Re[\epsilon_{L}], \Im[\epsilon_{L}], \Re[\epsilon_{R}], \Im[\epsilon_{R}],$ & \multirow{2}*{3.76/8} & \multirow{2}*{$-0.51(10)$} & \multirow{2}*{$-0.83(32)$} & \multirow{2}*{$0.016(100)$} & \multirow{2}*{$-0.028(320)$} & -- & -- & \multirow{2}*{$-0.007(400)$} & \multirow{2}*{$0.011(22)$} & \multirow{2}*{1.09} \\
& $\Re[\epsilon_{P}], \Im[\epsilon_{P}]$ & & & & & & & & & & \\
\hline
18 & $\Re[\epsilon_{L}], \Re[\epsilon_{R}], \Re[\epsilon_{P}]$ & 3.76/10 & $-0.031(30)$ & -- & $0.033(30)$ & -- & -- & -- & $-0.013(6)$ & -- & 1.09\\
\hline
\end{tabular}
}
\caption{Similar to Table~\ref{tab:electronfit}, but for the muon observables with $\chi^2_{SM}/dof = 4.94/14$, $\chi^2_{SM,S}/dof = 2.5/7$ and $\chi^2_{SM,P}/dof = 2.7/8$.}
\label{tab:muonfit}
\end{table}

\section{Deep Inelastic Neutrino-Nucleon Scattering}
\begin{figure}
\centering
    \includegraphics[scale=0.9]{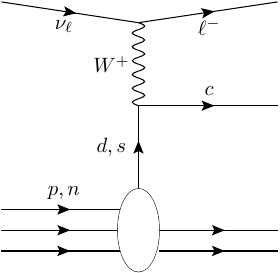}
\caption{Feynman diagram illustrating charged-current deep-inelastic scattering. This diagram with reversed arrows also applies to anti-neutrino interactions where an anti-strange quark ($\bar{s}$) is picked from the nucleon.}
\label{fig:DIS}
\end{figure}

In this section, we explore the sensitivity of SND@LHC to NP, as encapsulated in the WCs defined in Eq.~\eqref{eq:Heff}. Given the high energy of the incoming neutrinos, neutrino detection proceeds through charged-current deep-inelastic scattering (DIS):
\bea
\nu_{\ell}+ N \to \ell + X,
\label{proc-DIS}
\eea
where $N=p,n$ represents a nucleon, $\ell=e\,,\mu\,,\tau$ and $X$ denotes any hadron state. However, NP contributes specifically to  charm production, as shown in Fig.~\ref{fig:DIS}:
\bea
\nu_{\ell}+ N \to \ell + X_c,
\label{proc-DIS-c}
\eea
where $X_c$ denotes a possible charm hadron state. In this analysis, we evaluate the confidence level with which the WCs obtained by fitting low-energy observables, as summarized in Tables~\ref{tab:electronfit} and \ref{tab:muonfit}, can be probed at SND@LHC.
 
The sensitivity is determined through a $\chi^2$ analysis for each type of neutrino  using the following definition:
\beq
\chi^2\,=\,\sum_{i=1}^n\left(\frac{N_i^{\rm SM + NP}-N_i^{\rm SM}}{\sqrt{N_i^{\rm SM+NP}}}\right)^2 ,~~
\label{eq:chisq1}
\eeq
where $N_i^{\rm SM}$ are the number of differential events predicted by the SM and $N_i^{\rm SM+NP}$ are the total number of differential events, including contributions from NP, for the $i^{\rm th}$ bin in the energy distribution. 

We also assess the sensitivity to NP in the ratio, 
\beq
R^{}_{\mu e,i} \equiv \frac{N_{\mu,i}}{N_{e,i}} ,~~
\eeq
for WCs provided in Tables \ref{tab:electronfit} and \ref{tab:muonfit}, considering scenarios where NP effects are exclusively present in either muons or electrons. Here, $N^{}_{\mu(e),i}$ represents the model prediction of the number of events for the process $\nu_{\ell} + N \to \ell + X/X_c$ in the $i$-th energy bin. We then define the $\chi^2$ for this observable as
\beq
\chi^2 = \sum_{i=1}^n \left( \frac{R_{\mu e,i}^{\text{SM + NP}} - R_{\mu e,i}^{\text{SM}}}{\sigma_{R_{\mu e,i}^{\text{SM + NP}}}} \right)^2,~~
\label{eq:chisq2}
\eeq
where $\si_{R^{\rm SM+NP}_{\mu e,i}}$ is the statistical error in $R_{\mu e,i}$ and is calculated as follows.
\beq
\sigma_{R_{\mu e,i}}^2 \equiv R_{\mu e,i}^2 \left( \frac{\sigma_{e,i}^2}{N_{e,i}^2} + \frac{\sigma_{\mu,i}^2}{N_{\mu,i}^2} \right) = R_{\mu e,i}^2 \left( \frac{1}{N_{e,i}} + \frac{1}{N_{\mu,i}} \right),
\eeq
where in the final step we have assumed that the event distribution follows Poisson statistics so that  $\sigma^2_{\mu(e),i} = N_{\mu(e),i}$.

For signal or background, the total number of events is calculated from the DIS cross-section, $\sigma_{\nu N}$, as follows:
\begin{eqnarray}
N_i^{\rm model} &=& \mathcal{L} \times \sigma(pp \to \nu X) \times A_g \times P_{\rm int} (\sigma_{\nu N}^{\rm model}),\\
&=& \Phi_{i,\nu}\times P_{\rm int} (\sigma_{\nu N}^{\rm model})
\label{eq:Nevents}
\end{eqnarray}
where $\mathcal{L}=290$~fb$^{-1}$ is the $pp$ luminosity, $\sigma(pp \to \nu X)$ is the cross-section for neutrino production in $pp$ collisions, $A_g$ is the geometrical acceptance of SND@LHC, and the product of these factors constitutes the incoming neutrino spectra $\Phi_{i,\nu}$ in the $i^{\rm th}$ energy bin. $P_{\rm int} (\sigma_{\nu N}^{\rm model})$ represents the interaction probability of a neutrino with the detector, which is a function of the DIS cross-section. Here, model refers to background (SM) or signal (SM+NP).

\begin{figure}
\centering
    \includegraphics[scale=0.7]{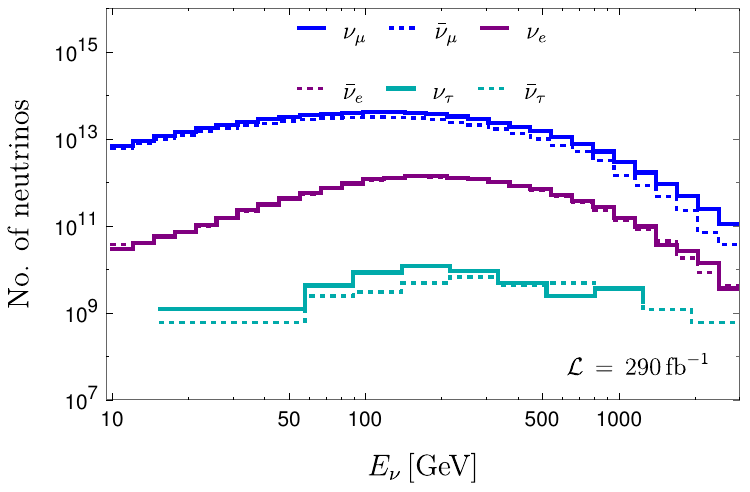}
\caption{Number of incoming neutrinos in SND@LHC as a function of energy. The spectrum is obtained from Ref.~\cite{Ahdida:2750060}, scaling to an expected integrated luminosity of $\mathcal{L}=290~\rm{fb}^{-1}$ and selecting only the neutrinos within the SND@LHC geometrical acceptance.}
\label{fig:neutrinospectra}
\end{figure}

We take and adapt the spectra of neutrinos produced at the ATLAS LHC interaction point and impinging onto the SND@LHC detector from Ref.~\cite{Ahdida:2750060}. These were obtained using the Pythia8, DPMJET and FLUKA library to simulate the production of neutrinos in $pp$ collisions, and their propagation through the machine elements until the location of the SND@LHC target. Only neutrinos within the SND@LHC geometrical acceptance $A_g$ are retained. These spectra are scaled up to an expected luminosity of $\mathcal{L}=290~\rm{fb}^{-1}$ and shown in Fig.~\ref{fig:neutrinospectra}.

To this end, the interaction probability of a neutrino with the detector is computed as:
\bea
P_{\rm int} &=\displaystyle \frac{{\rm A} \, \sigma_{\nu N}^{\rm model}}{\mathcal{S}} \frac{m_{\rm target}}{m_{\rm N}} \,
\eea
where $\sigma_{\nu N}^{\rm model}$ is the charge current scattering cross-section for Eqs.~\eqref{proc-DIS},~\eqref{proc-DIS-c}. The detector consists of 5 walls, each with 2×2 tungsten bricks with an area perpendicular to the beam direction of $\mathcal{S}=400\times400\,{\rm mm}^2$~\cite{Ahdida:2750060}. The total mass of the detector is $m_{\rm target}=830$ kg~\cite{Ahdida:2750060}, and $m_{\rm N}=3.05\times10^{-25}$~kg is the mass of a tungsten nucleus. The length of the detector is $l_{\rm target}=59\times5\,$mm, and the mass number of tungsten is A=183.

The cross-section $\sigma_{\nu N}$ is obtained as follows.

The differential cross-section for an incoming neutrino with energy $E_\nu$ scattering off a nucleon of mass $M$ in terms of the scattering amplitude is:
\beq
\frac{d\sigma_{\nu N}}{dxdy}=\frac{1}{32\pi M E_\nu}\int \frac{d\xi}{\xi}f(\xi) |\bar{\mathcal{M}}(\xi)|^2 \delta(\xi-x).
\eeq
where $\xi$ is the momentum fraction defined by $p_q^\mu = \xi p^\mu$, with $p_q^\mu$ being the four-momentum of the scattered quark and $p^\mu$ the target nucleon momentum. The function $f(\xi)$ represents the parton distribution function (PDF) within the nucleon.
We further decompose the differential cross-section of the neutrino-nucleon DIS into contributions from the SM, NP operators, and their interference terms: 
\beq
\frac{d\sigma_{\nu N}^{\rm SM+NP}}{dxdy}=\frac{d\sigma_{\nu N}^{\rm SM}}{dxdy}+\frac{d\sigma_{\nu N}^{\rm NP}}{dxdy}+\frac{d\sigma_{\nu N}^{\rm Interference}}{dxdy},
\label{eqnDiff}
\eeq
where $x$ is the Bjorken variable and $y$ is the inelasticity with $q$ being the four-momentum transfer of the leptonic probe and $\nu\,= -p\cdot q = M(E_\nu - E_\ell)$. The complete expressions for these terms are provided below for different interactions.

\textbf{Scalar interactions}:
Using Ref.~\cite{Liu:2015rqa}, we have 
\bea
\frac{d\sigma^{\rm SM}_{\nu N}}{dxdy}&=&\frac{G_F^2 M E_\nu}{\pi} \left(xy^2+\frac{ym_\ell^2}{2ME_\nu})F_1+ (1-y-\frac{Mxy}{2E_\nu}-\frac{m_\ell^2}{4E_\nu^2})F_2\right.\nn\\
&&\left. + (xy-\frac{xy^2}{2}-\frac{ym_\ell^2}{4ME_\nu})F_3 - \frac{m_\ell^2}{2ME_\nu} F_5 \right),\nonumber\\
\frac{d\sigma^{\rm S}_{\nu N}}{dxdy}&=&\frac{G_F^2 M E_\nu}{\pi}(\epsilon_S^2+\epsilon_P^2)y(xy+\frac{m_\ell^2}{2ME_\nu})F_1, \nonumber\\
\frac{d\sigma^{\rm SM,S}_{\nu N}}{dxdy}&=&0. 
\eea
The functions $F_i$ are given as
\bea
F_1 &=& \sum_{q,\bar{q}}f_{q,\bar{q}} (\xi,Q^2) V^2_{q,c},\nn\\
F_2 &=& 2\sum_{q,\bar{q}}\xi f_{q,\bar{q}} (\xi,Q^2) V^2_{q,c},\nn\\
F_3 &=& 2\sum_{q}f_{q} (\xi,Q^2) V^2_{q,c}-2\sum_{\bar{q}}f_{\bar{q}} (\xi,Q^2) V^2_{\bar{q},\bar{c}},\nn\\
F_5 &=& 2\sum_{q,\bar{q}}f_{q,\bar{q}} (\xi,Q^2) V^2_{q,c},
\eea
where $f_{q}$ and $f_{\bar{q}}$ are the parton distribution functions inside a nucleon, $V_{q,q'}$ is the CKM matrix element, and $Q^2=-q^2$.

\textbf{Vector and axial vector interactions}:
For vector and axial interactions, we have
\bea
\frac{d\sigma_{\nu N}^{\rm SM+VA}}{dxdy}\,&=&\,\frac{G_F^2M E_\nu}{\pi}\left[\left(\frac{|a'|^2+|b'|^2}{2}\right)\left(\frac{y m_\ell^2}{2M E_\nu}\, +\,xy^2\right)F_1(x) \right. \nl && \left. +\left(\frac{|a'|^2+|b'|}{2}^2\right)\left(1-y-\frac{xyM}{E_\nu}-\frac{m_\ell^2}{4E_\nu^2}\right)F_2(x) \right. \nl && \left. +{\rm Re}(a'b^{'*})\(xy-\frac{xy^2}{2}-\frac{ym_l^2}{4ME_\nu}\)F_3(x)-\left(\frac{|a'|^2+|b'|^2}{2}\right)\frac{m_l^2}{2ME_\nu}F_5(x)\right]\nl
\eea
where $a'\,=\,1+\epsilon_L+ \epsilon_R$ and $b'\,=1+ \epsilon_L- \epsilon_R$.

The total cross-section $\sigma_{\nu N}$ is obtained by integrating the differential cross-section over the following limits:
\bea
\frac{m_l^2}{2M(E_\nu-m_l)}\,\leq\,x\,&\leq&\,1,\\
A-B\,\leq\,y\,\leq\,A+B
\eea
where
\bea
A\,&=&\,\frac{1}{2}\left(1-\frac{m_l^2}{2ME_\nu x}-\frac{m_l^2}{2E_\nu^2}\right)\Big/\left(1+\frac{xM}{2E_\nu}\right),\\
B\,&=&\,\frac{1}{2}\left[\left(1-\frac{m_l^2}{2ME_\nu x}\right)^2-\frac{m_l^2}{E_\nu^2}\right]\Big/\left(1+\frac{xM}{2E_\nu}\right).
\eea

\section{Results}

The SND@LHC detector covers the pseudorapidity range of $7.2\,<\eta\,<\,8.4$~\cite{SNDLHC:2022ihg}. We obtain event distributions as functions of the neutrino energy, $E_\nu$, for both background (SM) and signal (SM + NP) events, by applying the appropriate pseudorapidity cuts. The NP scenarios we study are listed in Tables~\ref{tab:electronfit} and \ref{tab:muonfit} and were obtained by fitting to low-energy observables. We then use the energy distributions to calculate the sensitivity to NP at SND@LHC, by applying Eq.~\eqref{eq:Nevents} in conjunction with Eq.~\eqref{eq:chisq1} and Eq.~\eqref{eq:chisq2}. 

In principle, NP affecting charm quarks, as encoded in the effective Hamiltonian of Eq.~\eqref{eq:Heff}, will also impact the production flux of neutrinos. Ref.~\cite{Kling:2021gos} demonstrates that charm-hadron decays predominantly drive the electron-neutrino flux at energies above 200 GeV, while they contribute equally to the muon-neutrino flux, along with pion decays, at energies above 400 GeV. These effects have been thoroughly analyzed in Ref.~\cite{Falkowski:2021bkq} for all neutrino flavors. Their study shows that an NP pseudoscalar interaction with order-one coupling significantly amplifies the flux at production due to 2-body $D_s$ decays, increasing the electron-neutrino flux by up to $10^3 \times$  and the muon-neutrino flux by up to $27\times$ than in the SM. However, $\nu_e$ and $\nu_\mu$ production are dominated by three-body $D_s$ decays, which unlike two-body $D_s$ decays are not similarly enhanced. 
Therefore, these production and detection enhancements do not impact our analysis, as we only consider NP in the charm sector. We emphasize that, even if NP were present in first-generation quarks, stringent low-energy constraints impose strong limits on the corresponding Wilson coefficients, rendering current neutrino detectors ineffective in probing such NP operators. Furthermore, in our analysis even at a high luminosity, significant sensitivity to NP affecting muons is only observed when the final state is charm tagged. We, therefore, assume a 100\% charm-tagging efficiency.
\begin{figure}[!htbp]
\centering
    \begin{overpic}[width=0.49\textwidth]{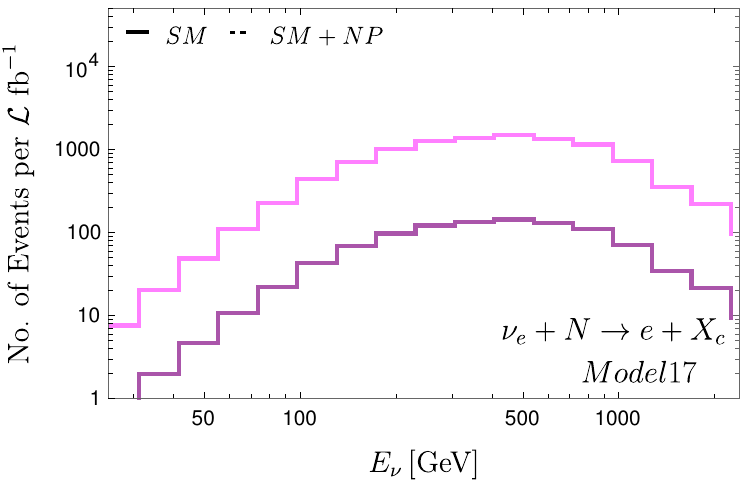}\put(57,39){\footnotesize 290~fb$^{-1}$}\put(57,51){\footnotesize 3000~fb$^{-1}$}
    \end{overpic}~~~~~
    \begin{overpic}[width=0.49\textwidth]{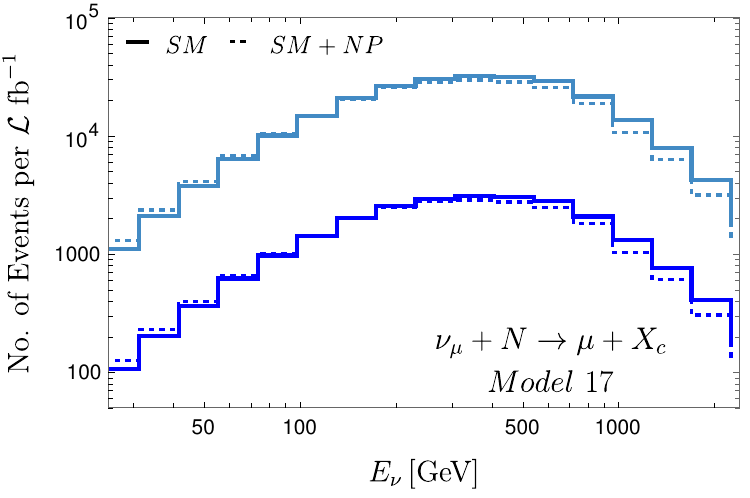}\put(52.5,41.5){\footnotesize 290~fb$^{-1}$}\put(56.5,51){\footnotesize 3000~fb$^{-1}$}\end{overpic}
\caption{Number of signal (SM + NP, dashed) and background (SM only, solid)  events as functions of the incoming neutrino energy ($E_\nu$) for $\nu_{e} + N \to e^- + X_c$ (left plot) and $\nu_{\mu} + N \to \mu + X_c$ (right plot.) The integrated luminosity $\cL$ has been set to the SND@LHC current projection of 290~fb$^{-1}$ for the lower (dark blue/dark pink) histograms, and to its high-lumi upgrade projection of 3000~fb$^{-1}$ for the upper (light blue/light pink) histograms. These results use couplings of model 17 from Tables~\ref{tab:electronfit} and \ref{tab:muonfit}, alongwith the pseudorapidity range  $7.2\,<\eta\,<8.4$ corresponding to the SND@LHC detector.}
\label{fig:muon_distribution}
\end{figure}

The resulting event distributions for the electron and muon neutrinos are presented in Fig.~\ref{fig:muon_distribution}. As can be seen in Fig.~\ref{fig:muon_distribution}, the NP contribution is negligible for the electron neutrino. In Table~\ref{tab:SensitivitySND_muon}, we present our results for the sensitivities of detecting NP with muons in the final state. This sensitivity is influenced by the left- and right-handed vector couplings. These results are based on the central values of NP couplings listed in Table~\ref{tab:muonfit}. Here, we only include those cases where the sensitivity exceeds $0.1\,\sigma$.
\begin{table}[!htbp]
\centering
\footnotesize
\begin{tabular}{|c|c|c|}\hline
Models & Sens($\nu_{\mu}+ N \to \mu + X_c$) & Sens$\left(\displaystyle \frac{\nu_{\mu}+ N \to \mu + X_c}{\nu_{e^-}+ N \to e^- + X_c}\right)$\\ \hline
1& 0.3$\sigma$& 0.1$\sigma$\\
17 & 0.4$\sigma$& 0.1$\sigma$ \\
\hline
\end{tabular}
\caption{Sensitivity (Sens) of detecting NP in the signal $\nu_{\mu} + N \to \mu + X_c$ at SND@LHC for models in Table \ref{tab:muonfit} where the significance exceeds 0.1$\sigma$. It also shows the sensitivity for finding NP in the ratio $\left(\frac{\nu_{\mu} + N \to \mu + X_c}{\nu_{e^-} + N \to e^- + X_c}\right)$, assuming NP is present only in the muon interaction (i.e., the numerator). The significances are calculated for a luminosity of 290 fb$^{-1}$.}
\label{tab:SensitivitySND_muon}
\end{table}
The high-luminosity upgrade of SND@LHC can collect ten times more data \cite{Abbaneo:2895224}, significantly enhancing the sensitivity to NP, even in models where no significant sensitivity is observed at $\mathcal{L}\,=\,290\,{\rm fb}^{-1}$. We assess the sensitivity for a luminosity of 3000 fb$^{-1}$ in Table~\ref{tab:SensitivitySND_LL}. Fig.~\ref{fig:muon_distribution} also includes event distributions for the high-lumi upgrade of SND@LHC.
\begin{table}[htp]
\centering
\footnotesize
\begin{tabular}{|c|c|c|}\hline
Models & Sens($\nu_{\mu}+ N \to \mu + X_c$) & Sens$\left(\displaystyle \frac{\nu_{\mu}+ N \to \mu + X_c}{\nu_{e^-}+ N \to e^- + X_c}\right)$ \\ \hline
1& 2.8$\sigma$&1.2$\sigma$ \\
2& 0.5$\sigma$&0.1$\sigma$\\
17 & 3.1$\sigma$&  1.4$\sigma$\\
\hline
\end{tabular}
\caption{Sensitivity of detecting NP in the signal $\nu_{\mu} + N \to \mu + X_c$ at SND@LHC, based on models listed in Table \ref{tab:muonfit}, assuming a luminosity of 3000 fb$^{-1}$.}
\label{tab:SensitivitySND_LL}
\end{table}

Finally, we note the importance of $\nu_\tau$ scattering at experiments such as SND@LHC. NP effective operators involving both the charm quark and the $\tau$ neutrino are difficult to constrain from low-energy measurements. Due to the relative closeness in the $\tau$ and charm masses, three-body decays are suppressed or prohibited due to phase-space effects. In these cases, neutrino scattering experiments can shed a significant amount of light on such NP operators. In this article, however, we restrict ourselves to the cases where neutrino scattering experiments can provide information complementary to low-energy constraints, and as such do not consider $\tau$-neutrino scattering.

\section{Conclusion}
In this paper, we have presented a detailed sensitivity analysis of new physics (NP) and lepton universality violation (LUV) in neutrino scattering at SND@LHC, focusing on processes involving a charm quark in the final state. Within the framework of effective field theory, NP effects were described using higher-dimensional four-Fermi operators. The Wilson coefficients for these dimension-6 operators were determined from fits to low-energy data, particularly from decays of charmed mesons and baryons. Using benchmark NP points derived from these low-energy analyses, we provide sensitivity estimates for detecting NP and LUV effects at SND@LHC. Both production and scattering processes of neutrinos were considered in evaluating NP effects.

Our study shows that SND@LHC exhibits significant sensitivity to these operators only when both charm tagging is effective and high luminosity is achieved, as NP couplings are tightly constrained by low-energy measurements. The inability of neutrino detectors to probe NP under these constraints aligns with the findings from FASER$\nu$, when their analyses of NP constraints and detector sensitivity are considered together~\cite{Falkowski:2021bkq}. However, effective charm tagging at high luminosity emerges as a key factor that could enhance the sensitivity to the $3\sigma$ level.  In principle, the high-lumi upgrade of SND@LHC could gather 10 times more statistics~\cite{Abbaneo:2895224}, which will lead to higher sensitivities to NP effects in the $\nu_\mu$ channel.

We also believe that a more comprehensive, multi-dimensional analysis could further improve the sensitivity. Additionally, our findings indicate that these detectors may be sensitive to NP operators that evade low-energy constraints but become relevant at high-energy scales. One such operator, as proposed in Ref.~\cite{Datta:1996gg}, fits this profile, though we leave a detailed investigation for future work. Furthermore, since effective couplings involving charmed particle decays with a final state $\nu_\tau$ receive less stringent constraints from low-energy experiments due to phase space restrictions, we note that neutrino scattering experiments can provide useful information about NP in such effective operators.

{\bf Acknowledgments}: This work was financially supported by the
U.S. National Science Foundation under Grant No.\ PHY-2310627 (BB) and PHY-2309937 (AD). BB acknowledges support in part from the NSF Grant No.\ PHY-2309135 to the Kavli Institute of Theoretical Physics (KITP) where part of this work was completed. DS is supported by funding from the European Union’s Horizon 2020 research and innovation programme under grant agreement No. 101002846 (ERC CoG “CosmoChart”). The work of EG is supported by the Swiss National Science Foundation (SNSF) under grant number 202065 (``Ambizione''). We are grateful to the SND@LHC collaboration for several useful discussions. We thank J.~Kopp, F.~Kling, and Z.~Tabrizi for pointing out that three-body decays play a vital role in the production of electron and muon neutrinos for SND@LHC.

\appendix

\section{Lepton Universality Violation in charged-current $B$ decays}
\label{app:LUVr}

Table \ref{tab:obs_meas} below summarizes experimental measurements and SM predictions in several charged-current $B$ decays. The $R^{\tau/\ell}_{D^{(*)}}$ values roughly show a 2.1-2.2 $\sigma$ deviation from the SM. However, these measurements are correlated and the combined deviation from the SM appears to be around 3.2$\sigma$ \cite{HFLAV:2022esi}. A 1.7$\sigma$ deviation is currently observed in $R_{J/\psi}^{\tau/\mu}$. While there is no observed deviation from the SM, current measurements allow LUV at the 3-5\% level in $R^{\mu/e}_{D^{(*)}}$. Together these observables provide strong hints of Lepton Universality Violating (LUV) new physics (NP) in charged-current $B$ decays.
\begin{table}[!htbp]
\begin{center}
\begin{tabular}{|c|c|c|} \hline\hline
Observable & SM Prediction & Measurement \\ \hline
$R_{D^*}^{\tau/\ell}$ & $0.254 \pm 0.005$ \cite{HFLAV:RDRDst2024update} & $0.287 \pm 0.012$ \cite{HFLAV:RDRDst2024update} \\
$R_{D}^{\tau/\ell}$ & $0.298 \pm 0.004$  \cite{HFLAV:RDRDst2024update} & $0.342 \pm 0.026$ \cite{HFLAV:RDRDst2024update} \\
$R_{J/\psi}^{\tau/\mu}$ & $0.283 \pm 0.048$ \cite{Watanabe:2017mip} & $0.71 \pm 0.17 \pm 0.18$ \cite{Aaij:2017tyk} \\
$R_{D^*}^{\mu/e}$ & $\sim 1.0$ & $1.04 \pm 0.05 \pm 0.01$ \cite{Belle:2017rcc} \\ \hline\hline
\end{tabular}
\end{center}
\caption{Standard Model predictions and measured values of observables that may provide hints of LUV NP in charged-current $B$ decays.}
\label{tab:obs_meas}
\end{table}

\section{Hadronic amplitudes for three body decays}
\label{app:hadamp}
The hadronic matrix element of the vector current for the three body decay $M_i \to M_j \ell \nu$ in terms of the form factors $f_{+,0}(q^2)$ is given as
\beq
\langle M_j (k^\prime)|\bar{q_j} \gamma_\mu c| M_i (k)\rangle = f_+(q^2)\((k+k^\prime)_\mu -\frac{m_{M_i}^2-m_{M_j}^2}{q^2}q_\mu\) + f_0(q^2)\frac{m_{M_i}^2-m_{M_j}^2}{q^2}q_\mu.
\eeq
Using the equation of motion, $i \partial_\mu(\bar q_j \gamma^\mu c) = (m_c - m_j) (\bar q_j c)$, 
 the scalar matrix element can be written as
\beq
\langle M_j |\bar{q_j} c| M_i\rangle =  f_0(q^2)\frac{m_{M_i}^2-m_{M_j}^2}{m_c - m_j}.
\eeq
The non-vanishing helicity amplitudes in terms of the form factors are given by
\bea
h_{V,0}^2(q^2) &=& \sqrt{\frac{\lambda(m_{M_i}^2, m_{M_j}^2, q^2)}{q^2}} f_+(q^2), \\
h_{V,t}^2(q^2) &=& \frac{m_{M_i}^2 - m_{M_j}^2}{\sqrt{q^2}} f_0(q^2),\\
h_{S}^2(q^2) &\simeq & \frac{m_{M_i}^2 - m_{M_j}^2}{m_c-m_j} f_0(q^2).
\eea

Similarly, the vector and axial vector hadronic matrix elements for the $M_i \to M_j^* \ell \nu$ decays as a function of the form factors can be written as
\bea 
\langle M^*(k,\varepsilon)|\bar q_j\gamma_\mu c|\overline{M}_i(p)\rangle &=& -i\epsilon_{\mu\nu\rho\sigma}\varepsilon^{\nu *}p^\rho k^\sigma \frac{2V(q^2)}{m_{M_i}+m_{M_j^*}}, \\
\langle M^*(k,\varepsilon)|\bar q_j\gamma_\mu \gamma_5 c|\overline{M}_i(p)\rangle &=& \varepsilon^{\mu *}(m_{M_i}+m_{M_{j^*}})A_1(q^2) - (p+k)_\mu (\varepsilon^* q)\frac{A_2(q^2)}{m_{M_i}+m_{M_{j^*}}} \nl
&& ~~-q_\mu (\varepsilon^* q)\frac{2m_{M_j^*}}{q^2}(A_3(q^2)-A_0(q^2)),
\eea 
where
\beq 
A_3(q^2) = \frac{m_{M_i}+m_{M_{j^*}}}{2m_{M_j^*}}A_1(q^2) - \frac{m_{M_i}-m_{M_{j^*}}}{2m_{M_j^*}}A_2(q^2).
\eeq 
The kinematic constraint at zero recoil gives $A_3(0)=A_0(0)$. The pseudoscalar matrix element is determined using the equation of motion $i \partial_\mu (\bar q_j \gamma^\mu\gamma^5 c) = -(m_c + m_j)(\bar q_j \gamma^5 c)$ and is given by
\beq
\langle M^*(k,\varepsilon)|\bar q_j \gamma_5 c|\overline{M}_i(p)\rangle = -(\varepsilon^* q)\frac{2m_{M_{j^*}}}{m_c + m_j}A_0(q^2).
\eeq 
 The helicity amplitudes in terms of the form factors are given as 
 \bea
 H_{V,\pm}(q^2) &=& (m_{M_i}+m_{M_{j^*}})A_1(q^2) \mp \frac{\sqrt{\lambda(m_{M_i}^2,m_{M_j^*}^2,q^2)}}{m_{M_i}+m_{M_j^*}}V(q^2),\\
 H_{V,0}(q^2) &=& \frac{m_{M_i}+m_{M_{j^*}}}{2m_{M_j^*}\sqrt{q^2}}\bigg \{ -(m_{M_i}^2-m_{M_j^*}^2-q^2)A_1(q^2) \nl 
 && \hspace{3cm}+\frac{\lambda(m_{M_i}^2,m_{M_j^*}^2,q^2)}{(m_{M_i}+m_{M_j^*})^2}A_2(q^2) \bigg \}, \\
 H_{V,t}(q^2) &=& - \sqrt{\frac{\lambda(m_{M_i}^2,m_{M_j^*}^2,q^2)}{q^2}}A_0(q^2),\\
H_{S}(q^2) &\simeq & - \frac{\sqrt{\lambda(m_{M_i}^2,m_{M_j^*}^2,q^2)}}{m_c+m_j}A_0(q^2).
 \eea 

As mentioned earlier, we incorporate the hadronic matrix elements for the baryonic $\Lambda_c \to \Lambda \ell \nu$ decays from Ref.~\cite{Datta:2017aue}.

\section{Numerical inputs and form factors}
\label{app:formfactors}
Here we collect the relevant numerical inputs (in Table~\ref{tab:input}) and form factors for the charmed hadron decays to pions, kaons, and other light pseudoscalar and vector resonances used in the analysis. 

\begin{table}[htp!]
    \centering
    \begin{tabular}{|c|c|c|c|}
    \hline
         $G_F$ & $1.166\times 10^{-5}$~\cite{ParticleDataGroup:2022pth}& $m_{K^+}$ & $493.677$~MeV~\cite{ParticleDataGroup:2022pth}\\
         $f_{D_s}$ & $249.9(5)$~MeV ~\cite{FlavourLatticeAveragingGroupFLAG:2021npn}& $m_{K^0}$ & $497.611$~MeV~\cite{ParticleDataGroup:2022pth} \\
         $f_D$ & $212.0(7)$~MeV~\cite{FlavourLatticeAveragingGroupFLAG:2021npn} & $m_{\pi^+}$ & $139.57$~MeV~\cite{ParticleDataGroup:2022pth}  \\
         $|V_{cs}|$ & $0.983(2)(18)$~\cite{FlavourLatticeAveragingGroupFLAG:2021npn} & $m_{\pi^0}$ & $134.98$~MeV~\cite{ParticleDataGroup:2022pth}\\
         $|V_{cd}|$ & $0.2179(7)(57)$~\cite{FlavourLatticeAveragingGroupFLAG:2021npn} & $m_{K^{*+}}$ & $891.67$~MeV~\cite{ParticleDataGroup:2022pth} \\
         $\tau_{D_s}$ & $0.501$~ps~\cite{ParticleDataGroup:2022pth} & $m_{K^{*0}}$ & $895.55$~MeV~\cite{ParticleDataGroup:2022pth} \\
         $\tau_{D^+}$ & $1.033$~ps~\cite{ParticleDataGroup:2022pth} & $m_{\rho^+}$ & $766.5$~MeV~\cite{ParticleDataGroup:2022pth} \\
         $\tau_{D^0}$ & $0.410$~ps~\cite{ParticleDataGroup:2022pth} & $m_{\rho^0}$ & $775.26$~MeV~\cite{ParticleDataGroup:2022pth} \\
         $m_{D_s^+}$ & $1968.35$~MeV~\cite{FlavourLatticeAveragingGroupFLAG:2021npn} & $m_\phi$ & $1.019$~GeV~\cite{ParticleDataGroup:2022pth} \\
         $m_{D^+}$ & $1869.66$~MeV~\cite{ParticleDataGroup:2022pth} &  $m_\eta$ & $547.86$~MeV~\cite{ParticleDataGroup:2022pth} \\
         $m_{D^0}$ & $1864.84$~MeV~\cite{ParticleDataGroup:2022pth} & $m_{\eta^\prime}$ & $957.78$~Me~\cite{ParticleDataGroup:2022pth} \\
         $m_{D^*}$ & $2.1122$~GeV~\cite{ParticleDataGroup:2022pth} & $m_{\Lambda_c}$ & $2.286$~GeV~\cite{ParticleDataGroup:2022pth} \\
         $m_{D^{*0}}$ & $2.317$~GeV~\cite{ParticleDataGroup:2022pth} & $m_{\Lambda}$ & $1.116$~GeV~\cite{ParticleDataGroup:2022pth} \\
         \hline
    \end{tabular}
    \caption{List of all numerical inputs used in the analysis.}
    \label{tab:input}
\end{table}

For the $D\to\pi\ell\nu_\ell$ transition, the $q^2$ dependence of two relevant form factors $f_{+,0}^{D\to\pi}$ can be expressed as \cite{Lubicz:2017syv}:
\bea
f_+^{D\to\pi}(q^2) &=& \frac{1}{1-P_V q^2}\bigg\{f^{D\to\pi}(0)+c_+^{D\to\pi}(z-z_0)\(1+\frac{z+z_0}{2}\)\bigg\},\,\\
f_0^{D\to\pi}(q^2) &=& \frac{1}{1-P_S q^2}\bigg\{f^{D\to\pi}(0)+c_0^{D\to\pi}(z-z_0)\(1+\frac{z+z_0}{2}\)\bigg\},\,
\eea 
where $z$ is defined as
\beq
z = \frac{\sqrt{t_+-q^2}-\sqrt{t_+-t_0}}{\sqrt{t_+-q^2}+\sqrt{t_+-t_0}},
\label{eq:z-form}
\eeq 
with
\bea
t_+ &=& (m_D+m_\pi)^2,\label{eq:tplus-form}\\
t_0 &=& (m_D+m_\pi)(\sqrt{m_D}-\sqrt{m_\pi})^2.
\label{eq:t0-form}
\eea
The other parameters that appear in the z-expansion of the form factors are collected below in Table~\ref{tab:FFpi}.
\begin{table}[htp!]
    \centering
    \begin{tabular}{|c|c|c|c|c|}
    \hline
         $f^{D\to\pi}(0)$ & $c_+^{D\to\pi}$ & $P_V~({\rm GeV}^{-2})$ & $c_0^{D\to\pi}$ & $P_S~({\rm GeV}^{-2})$   \\
         \hline
         ~0.6117(354)~ & ~-1.985(347)~ & ~0.1314(127)~ & ~-1.188(256)~ & ~0.0342(122)~\\ 
         \hline
    \end{tabular}
    \caption{Values of the parameters taken from \cite{Lubicz:2017syv} that appear in the z-expansion of the vector and scalar form factors of the $D \to \pi$ transition.}
    \label{tab:FFpi}
\end{table} 

A similar kind of z-expansion is also employed in case of the $D\to K$ transition are expressed as
\bea
f_+^{D\to K}(q^2) &=& \frac{1}{1-q^2/m_{D^*}^2}\(a_0^+ + a_1^+(z-z^3/3)+a_2^+(z^2 +2z^3/3)\),\\
f_0^{D\to K}(q^2) &=& \frac{1}{1-q^2/m_{D^{*0}}^2}\(a_0^0 + a_1^0~z + a_2^0~z^2)\),
\eea 
with $m_\pi \leftrightarrow m_K$ in Eqs.~\eqref{eq:tplus-form}-\eqref{eq:t0-form}. The kinematic constraint at zero recoil $f_+^{D\to K}(0) = f_0^{D\to K}(0)$ reduces one parameter out of the six while rest of the coefficients are taken from the most recent FLAG $N_f = 2 + 1 +1$ computation of the form factors \cite{FlavourLatticeAveragingGroupFLAG:2021npn} as listed in Table~\ref{tab:FFK}. 

\begin{table}[htp!]
    \centering
    \begin{tabular}{|c|c|c|c|c|}
    \hline
    $a_0^+$ & $a_1^+$ & $a_2^+$ & $a_0^0$ & $a_1^0$  \\
    \hline
    ~0.7877(87)~ & ~-0.97(18)~ & ~-0.3(2.0)~ & ~0.6959(47)~ & ~0.775(69)~ \\
    \hline
    \end{tabular}
    \caption{Values of form factor parameters relevant to the $D\to K$ transition as obtained by \cite{FlavourLatticeAveragingGroupFLAG:2021npn}.}
    \label{tab:FFK}
\end{table}

The form factors for the $D_s \to (\eta,\eta^\prime)\ell\nu$ transition are calculated based on the light cone QCD sum rules (LCSR) as in Ref.~\cite{Azizi:2010zj}. The $q^2$ dependence of the vector form factors is given by the form
\beq
f_\pm^{D_s\to\eta}(q^2) = \frac{f_\pm(0)^{D_s\to\eta}}{1-\alpha_\pm \hat{q}+\beta_\pm \hat{q}^2},
\eeq 
with $\hat{q} = q^2/m_{D_s}^2$ and the parameters $\alpha_\pm$, $\beta_\pm$ and $f_\pm(0)^{D_s\to\eta}$ are listed in Table.~\ref{tab:FFEta}. The scalar form factor $f_0^{D_s\to\eta}$ is related to the vector ones by the relation
\beq
f_0^{D_s\to\eta}(q^2) = f_+^{D_s\to\eta}(q^2) + \frac{q^2}{m_{D_s}^2-m_\eta^2}f_-^{D_s\to\eta}(q^2).
\eeq
The form factors $f_i^{D_s\to \eta^\prime}$ are related to $f_i^{D_s\to \eta}$ as \cite{Azizi:2010zj}
\beq 
\frac{|f_i^{D_s\to \eta}|}{|f_i^{D_s\to \eta^\prime}|} = \tan \varphi,
\eeq 
where $\varphi = (39.7 \pm 0.7)^\circ$~\cite{Azizi:2010zj} is the mixing angle between the $\eta$ and $\eta^\prime$ states.

\begin{table}[htp!]
    \centering
    \begin{tabular}{|c|c|c|c|c|c|}
    \hline
    $f_+^{D_s\to \eta}(0)$ & $\alpha_+$ & $\beta_+$ & $f_-^{D_s\to \eta}(0)$ & $\alpha_-$ & $\beta_-$  \\
    \hline
    ~0.45(14)~ & ~1.96(63)~ & ~1.12(36)~ & ~-0.44(13)~ & ~2.05(65)~ & ~1.08(35)~\\
    \hline
    \end{tabular}
    \caption{Values of form factor parameters relevant to the $D_s\to \eta$ transition as obtained by the LCSR calculation in Ref.~\cite{Azizi:2010zj}.}
    \label{tab:FFEta}
\end{table}

A simple single pole parametrization is employed to calculate the form factors for the pseudoscalar $P = D, D_s$ to vector $M = K^*, \phi, \rho$ meson transitions given as
\beq
F_i^M(q^2) = \frac{F_i^M(0)}{1-q^2/m_{R,i}^2},
\eeq 
where $F_i = A_1, A_2, A_0, V$ and $m_{R,i}$ are the respective pole masses. The kinematic constraint $A_0(0) = A_3(0) =( \frac{m_P + m_M}{2m_M}A_1(0) - \frac{m_P - m_M}{2m_M}A_2(0)) $ relates the parameters at zero recoil. We list the values of the relevant parameters in Table.~\ref{tab:FFvec}.

\begin{table}[htp!]
    \centering
    \begin{tabular}{|c|c|c|c|c|c|c|c|}
    \hline
    \multirow{2}{*}{$F_i$} & \multicolumn{2}{c|}{$K^*$~\cite{BaBar:2010vmf}} & \multicolumn{2}{c|}{$\rho$~\cite{Fu:2018yin}} &\multicolumn{2}{c|}{$\phi$~\cite{Donald:2013pea}} \\
    \cline{2-7}
    & $F_i(0)$ & $m_{R,i}$~(GeV) & $F_i(0)$ & $m_{R,i}$~(GeV) & $F_i(0)$ & $m_{R,i}$~(GeV) \\
    \hline
    ~$A_1$~ & ~0.620(85)~ & ~2.459~ & ~$0.58^{+0.065}_{-0.050}$~ & ~2.427~ & ~0.615(24)~ & ~2.459~\\
    ~\multirow{2}{*}{$A_2$}~ & ~$r_2 A_1(0)$~ & \multirow{2}{*}{2.459} & \multirow{2}{*}{$0.468^{+0.052}_{-0.053}$} & \multirow{2}{*}{2.427} & \multirow{2}{*}{0.457(78)} & \multirow{2}{*}{2.459}\\
    & $r_2 = 0.801(30)$ & & & & &\\ 
    ~\multirow{2}{*}{$V$}~ & ~$r_V A_1(0)$~ & \multirow{2}{*}{2.112} & \multirow{2}{*}{$0.815^{+0.070}_{-0.051}$} & \multirow{2}{*}{2.007} & \multirow{2}{*}{1.059(124)} & \multirow{2}{*}{2.112}\\
    & $r_V = 1.463(35)$ & & & & &\\
    \hline
    \end{tabular}
    \caption{Values of form factor parameters used in our analysis for the $P \to M$ transitions along with the reference from which they are taken.}
    \label{tab:FFvec}
\end{table}

For the semileptonic baryonic decay, we employ the most recent lattice QCD computation of the form factors reported in~\cite{Meinel:2016dqj}. There are a total of six form factors $f_+, f_\perp, f_0, g_+, g_\perp, g_0$ which are parameterized using a simple pole z-expansion of the form
\beq
f(q^2) = \frac{1}{1-q^2/(m_{\rm pole}^f)^2} \sum_{n=0}^{n_{\rm max}} a_n^f\[z(q^2)\]^n,
\eeq
where z is defined as in Eq.~\eqref{eq:z-form} with $t_0 = q^2_{\rm max} = (m_{\Lambda_c}-m_\Lambda)^2$ and $t_+ = (m_D + m_K)^2$. The pole masses are $m_{\rm pole}^{f_+,f_\perp} = 2.112$~GeV, $m_{\rm pole}^{f_0} = 2.318$~GeV, $m_{\rm pole}^{g_+,g_\perp} = 2.460$~GeV and $m_{\rm pole}^{g_0} = 1.968$~GeV. The ``nominal fit" results for the form factor parameters provided in ~\cite{Meinel:2016dqj} have been used for our analysis. 

\bibliographystyle{JHEP}
\bibliography{references}

\providecommand{\href}[2]{#2}\begingroup\raggedright\begin{thebibliography}{10}

\bibitem{HFLAV:2022esi}
{\scshape HFLAV} collaboration, \emph{{Averages of b-hadron, c-hadron, and
  $\tau$-lepton properties as of 2021}},
  \href{https://doi.org/10.1103/PhysRevD.107.052008}{\emph{Phys. Rev. D}
  {\bfseries 107} (2023) 052008}
  [\href{https://arxiv.org/abs/2206.07501}{{\ttfamily 2206.07501}}].

\bibitem{Lees:2012xj}
{\scshape BaBar} collaboration, \emph{{Evidence for an excess of $\bar{B} \to
  D^{(*)} \tau^-\bar{\nu}_\tau$ decays}},
  \href{https://doi.org/10.1103/PhysRevLett.109.101802}{\emph{Phys. Rev. Lett.}
  {\bfseries 109} (2012) 101802}
  [\href{https://arxiv.org/abs/1205.5442}{{\ttfamily 1205.5442}}].

\bibitem{Lees:2013uzd}
{\scshape BaBar} collaboration, \emph{{Measurement of an Excess of $\bar{B} \to
  D^{(*)}\tau^- \bar{\nu}_\tau$ Decays and Implications for Charged Higgs
  Bosons}}, \href{https://doi.org/10.1103/PhysRevD.88.072012}{\emph{Phys. Rev.
  D} {\bfseries 88} (2013) 072012}
  [\href{https://arxiv.org/abs/1303.0571}{{\ttfamily 1303.0571}}].

\bibitem{Aaij:2015yra}
{\scshape LHCb} collaboration, \emph{{Measurement of the ratio of branching
  fractions $\mathcal{B}(\bar{B}^0 \to
  D^{*+}\tau^{-}\bar{\nu}_{\tau})/\mathcal{B}(\bar{B}^0 \to
  D^{*+}\mu^{-}\bar{\nu}_{\mu})$}},
  \href{https://doi.org/10.1103/PhysRevLett.115.111803}{\emph{Phys. Rev. Lett.}
  {\bfseries 115} (2015) 111803}
  [\href{https://arxiv.org/abs/1506.08614}{{\ttfamily 1506.08614}}].

\bibitem{Huschle:2015rga}
{\scshape Belle} collaboration, \emph{{Measurement of the branching ratio of
  $\bar{B} \to D^{(\ast)} \tau^- \bar{\nu}_\tau$ relative to $\bar{B} \to
  D^{(\ast)} \ell^- \bar{\nu}_\ell$ decays with hadronic tagging at Belle}},
  \href{https://doi.org/10.1103/PhysRevD.92.072014}{\emph{Phys. Rev. D}
  {\bfseries 92} (2015) 072014}
  [\href{https://arxiv.org/abs/1507.03233}{{\ttfamily 1507.03233}}].

\bibitem{Sato:2016svk}
{\scshape Belle} collaboration, \emph{{Measurement of the branching ratio of
  $\bar{B}^0 \rightarrow D^{*+} \tau^- \bar{\nu}_{\tau}$ relative to $\bar{B}^0
  \rightarrow D^{*+} \ell^- \bar{\nu}_{\ell}$ decays with a semileptonic
  tagging method}},
  \href{https://doi.org/10.1103/PhysRevD.94.072007}{\emph{Phys. Rev. D}
  {\bfseries 94} (2016) 072007}
  [\href{https://arxiv.org/abs/1607.07923}{{\ttfamily 1607.07923}}].

\bibitem{Hirose:2016wfn}
{\scshape Belle} collaboration, \emph{{Measurement of the $\tau$ lepton
  polarization and $R(D^*)$ in the decay $\bar{B} \to D^* \tau^-
  \bar{\nu}_\tau$}},
  \href{https://doi.org/10.1103/PhysRevLett.118.211801}{\emph{Phys. Rev. Lett.}
  {\bfseries 118} (2017) 211801}
  [\href{https://arxiv.org/abs/1612.00529}{{\ttfamily 1612.00529}}].

\bibitem{Aaij:2017uff}
{\scshape LHCb} collaboration, \emph{{Measurement of the ratio of the $B^0 \to
  D^{*-} \tau^+ \nu_{\tau}$ and $B^0 \to D^{*-} \mu^+ \nu_{\mu}$ branching
  fractions using three-prong $\tau$-lepton decays}},
  \href{https://doi.org/10.1103/PhysRevLett.120.171802}{\emph{Phys. Rev. Lett.}
  {\bfseries 120} (2018) 171802}
  [\href{https://arxiv.org/abs/1708.08856}{{\ttfamily 1708.08856}}].

\bibitem{Hirose:2017dxl}
{\scshape Belle} collaboration, \emph{{Measurement of the $\tau$ lepton
  polarization and $R(D^*)$ in the decay $\bar{B} \rightarrow D^* \tau^-
  \bar{\nu}_\tau$ with one-prong hadronic $\tau$ decays at Belle}},
  \href{https://doi.org/10.1103/PhysRevD.97.012004}{\emph{Phys. Rev. D}
  {\bfseries 97} (2018) 012004}
  [\href{https://arxiv.org/abs/1709.00129}{{\ttfamily 1709.00129}}].

\bibitem{Aaij:2017deq}
{\scshape LHCb} collaboration, \emph{{Test of Lepton Flavor Universality by the
  measurement of the $B^0 \to D^{*-} \tau^+ \nu_{\tau}$ branching fraction
  using three-prong $\tau$ decays}},
  \href{https://doi.org/10.1103/PhysRevD.97.072013}{\emph{Phys. Rev. D}
  {\bfseries 97} (2018) 072013}
  [\href{https://arxiv.org/abs/1711.02505}{{\ttfamily 1711.02505}}].

\bibitem{Belle:2019gij}
{\scshape Belle} collaboration, \emph{{Measurement of $\mathcal{R}(D)$ and
  $\mathcal{R}(D^{*})$ with a semileptonic tagging method}},
  \href{https://arxiv.org/abs/1904.08794}{{\ttfamily 1904.08794}}.

\bibitem{LHCb:2023zxo}
{\scshape LHCb} collaboration, \emph{{Measurement of the ratios of branching
  fractions $\mathcal{R}(D^{*})$ and $\mathcal{R}(D^{0})$}},
  \href{https://doi.org/10.1103/PhysRevLett.131.111802}{\emph{Phys. Rev. Lett.}
  {\bfseries 131} (2023) 111802}
  [\href{https://arxiv.org/abs/2302.02886}{{\ttfamily 2302.02886}}].

\bibitem{LHCb:2023uiv}
{\scshape LHCb} collaboration, \emph{{Test of lepton flavor universality using
  $B^0\to D^{*-}\tau^+\nu_\tau$ decays with hadronic $\tau$ channels}},
  \href{https://doi.org/10.1103/PhysRevD.108.012018}{\emph{Phys. Rev. D}
  {\bfseries 108} (2023) 012018}
  [\href{https://arxiv.org/abs/2305.01463}{{\ttfamily 2305.01463}}].

\bibitem{Belle-II:2024ami}
{\scshape Belle-II} collaboration, \emph{{A test of lepton flavor universality
  with a measurement of $R(D^{*})$ using hadronic $B$ tagging at the Belle II
  experiment}},  \href{https://arxiv.org/abs/2401.02840}{{\ttfamily
  2401.02840}}.

\bibitem{LHCb:2024jll}
{\scshape LHCb} collaboration, \emph{{Measurement of the branching fraction
  ratios $R(D^{+})$ and $R(D^{*+})$ using muonic $\tau$ decays}},
  \href{https://arxiv.org/abs/2406.03387}{{\ttfamily 2406.03387}}.

\bibitem{Aaij:2017tyk}
{\scshape LHCb} collaboration, \emph{{Measurement of the ratio of branching
  fractions
  $\mathcal{B}(B_c^+\,\to\,J/\psi\tau^+\nu_\tau)$/$\mathcal{B}(B_c^+\,\to\,J/\psi\mu^+\nu_\mu)$}},
  \href{https://doi.org/10.1103/PhysRevLett.120.121801}{\emph{Phys. Rev. Lett.}
  {\bfseries 120} (2018) 121801}
  [\href{https://arxiv.org/abs/1711.05623}{{\ttfamily 1711.05623}}].

\bibitem{HFLAV:RDRDst2024update}
{\scshape HFLAV} collaboration, Y.S.~Amhis et~al., ``{Preliminary average of
  $R(D)$ and $R(D^*)$ for Moriond 2024}.''
  \url{https://hflav-eos.web.cern.ch/hflav-eos/semi/moriond24/html/RDsDsstar/RDRDs.html},
  2024.

\bibitem{Belle-II:2023esi}
{\scshape Belle-II} collaboration, \emph{{Evidence for $B^+\to K^+\nu{\bar\nu}$
  decays}}, \href{https://doi.org/10.1103/PhysRevD.109.112006}{\emph{Phys. Rev.
  D} {\bfseries 109} (2024) 112006}
  [\href{https://arxiv.org/abs/2311.14647}{{\ttfamily 2311.14647}}].

\bibitem{Parrott:2022zte}
{\scshape HPQCD} collaboration, \emph{{Standard Model predictions for $B\to
  K\ell^+\ell^-$, $B\to K\ell_1^- \ell_2^+$ and $B\to K\nu\bar{\nu}$ using form
  factors from $N_f=2+1+1$ lattice QCD}},
  \href{https://doi.org/10.1103/PhysRevD.107.014511}{\emph{Phys. Rev. D}
  {\bfseries 107} (2023) 014511}
  [\href{https://arxiv.org/abs/2207.13371}{{\ttfamily 2207.13371}}].

\bibitem{LHCb:2022qnv}
{\scshape LHCb} collaboration, \emph{{Test of lepton universality in $b \to s
  \ell^+ \ell^-$ decays}},
  \href{https://doi.org/10.1103/PhysRevLett.131.051803}{\emph{Phys. Rev. Lett.}
  {\bfseries 131} (2023) 051803}
  [\href{https://arxiv.org/abs/2212.09152}{{\ttfamily 2212.09152}}].

\bibitem{Capdevila:2023yhq}
B.~Capdevila, A.~Crivellin and J.~Matias, \emph{{Review of semileptonic $B$
  anomalies}},
  \href{https://doi.org/10.1140/epjs/s11734-023-01012-2}{\emph{Eur. Phys. J.
  ST} {\bfseries 1} (2023) 20}
  [\href{https://arxiv.org/abs/2309.01311}{{\ttfamily 2309.01311}}].

\bibitem{Bhattacharya:2016mcc}
B.~Bhattacharya, A.~Datta, J.-P.~Gu\'evin, D.~London and R.~Watanabe,
  \emph{{Simultaneous Explanation of the $R_K$ and $R_{D^{(*)}}$ Puzzles: a
  Model Analysis}}, \href{https://doi.org/10.1007/JHEP01(2017)015}{\emph{JHEP}
  {\bfseries 01} (2017) 015}
  [\href{https://arxiv.org/abs/1609.09078}{{\ttfamily 1609.09078}}].

\bibitem{Bhattacharya:2014wla}
B.~Bhattacharya, A.~Datta, D.~London and S.~Shivashankara, \emph{{Simultaneous
  Explanation of the $R_K$ and $R(D^{(*)})$ Puzzles}},
  \href{https://doi.org/10.1016/j.physletb.2015.02.011}{\emph{Phys. Lett. B}
  {\bfseries 742} (2015) 370}
  [\href{https://arxiv.org/abs/1412.7164}{{\ttfamily 1412.7164}}].

\bibitem{Rashed:2012bd}
A.~Rashed, M.~Duraisamy and A.~Datta, \emph{{Nonstandard interactions of tau
  neutrino via charged Higgs and $W'$ contribution}},
  \href{https://doi.org/10.1103/PhysRevD.87.013002}{\emph{Phys. Rev. D}
  {\bfseries 87} (2013) 013002}
  [\href{https://arxiv.org/abs/1204.2023}{{\ttfamily 1204.2023}}].

\bibitem{Rashed:2013dba}
A.~Rashed, P.~Sharma and A.~Datta, \emph{{Tau neutrino as a probe of
  nonstandard interaction}},
  \href{https://doi.org/10.1016/j.nuclphysb.2013.10.022}{\emph{Nucl. Phys. B}
  {\bfseries 877} (2013) 662}
  [\href{https://arxiv.org/abs/1303.4332}{{\ttfamily 1303.4332}}].

\bibitem{Liu:2015rqa}
H.~Liu, A.~Rashed and A.~Datta, \emph{{Probing lepton nonuniversality in tau
  neutrino scattering}},
  \href{https://doi.org/10.1103/PhysRevD.93.039902}{\emph{Phys. Rev. D}
  {\bfseries 92} (2015) 073016}
  [\href{https://arxiv.org/abs/1505.04594}{{\ttfamily 1505.04594}}].

\bibitem{FASER:2019dxq}
{\scshape FASER} collaboration, \emph{{Detecting and Studying High-Energy
  Collider Neutrinos with FASER at the LHC}},
  \href{https://doi.org/10.1140/epjc/s10052-020-7631-5}{\emph{Eur. Phys. J. C}
  {\bfseries 80} (2020) 61} [\href{https://arxiv.org/abs/1908.02310}{{\ttfamily
  1908.02310}}].

\bibitem{SNDLHC:2022ihg}
{\scshape SND@LHC} collaboration, \emph{{SND@LHC: the scattering and neutrino
  detector at the LHC}},
  \href{https://doi.org/10.1088/1748-0221/19/05/P05067}{\emph{JINST} {\bfseries
  19} (2024) P05067} [\href{https://arxiv.org/abs/2210.02784}{{\ttfamily
  2210.02784}}].

\bibitem{Falkowski:2021bkq}
A.~Falkowski, M.~Gonz\'alez-Alonso, J.~Kopp, Y.~Soreq and Z.~Tabrizi,
  \emph{{EFT at FASER$\nu$}},
  \href{https://doi.org/10.1007/JHEP10(2021)086}{\emph{JHEP} {\bfseries 10}
  (2021) 086} [\href{https://arxiv.org/abs/2105.12136}{{\ttfamily
  2105.12136}}].

\bibitem{Ahdida:2750060}
C.~Ahdida et~al., \emph{{SND@LHC - Scattering and Neutrino Detector at the
  LHC}},  Tech. Rep.
  \href{https://cds.cern.ch/record/2750060}{CERN-LHCC-2021-003, LHCC-P-016},
  CERN, Geneva (2021).

\bibitem{Becirevic:2020rzi}
D.~Be\v{c}irevi\'c, F.~Jaffredo, A.~Pe\~nuelas and O.~Sumensari, \emph{{New
  Physics effects in leptonic and semileptonic decays}},
  \href{https://doi.org/10.1007/JHEP05(2021)175}{\emph{JHEP} {\bfseries 05}
  (2021) 175} [\href{https://arxiv.org/abs/2012.09872}{{\ttfamily
  2012.09872}}].

\bibitem{Kopp:2024yvh}
J.~Kopp, N.~Rocco and Z.~Tabrizi, \emph{{Unleashing the Power of EFT in
  Neutrino-Nucleus Scattering}},
  \href{https://arxiv.org/abs/2401.07902}{{\ttfamily 2401.07902}}.

\bibitem{Buchmuller:1986zs}
W.~Buchmuller, R.~Ruckl and D.~Wyler, \emph{{Leptoquarks in Lepton - Quark
  Collisions}}, \href{https://doi.org/10.1016/0370-2693(87)90637-X}{\emph{Phys.
  Lett. B} {\bfseries 191} (1987) 442}.

\bibitem{Sakaki:2013bfa}
Y.~Sakaki, M.~Tanaka, A.~Tayduganov and R.~Watanabe, \emph{{Testing leptoquark
  models in $\bar B \to D^{(*)} \tau \bar\nu$}},
  \href{https://doi.org/10.1103/PhysRevD.88.094012}{\emph{Phys. Rev. D}
  {\bfseries 88} (2013) 094012}
  [\href{https://arxiv.org/abs/1309.0301}{{\ttfamily 1309.0301}}].

\bibitem{Bhattacharya:2022bdk}
B.~Bhattacharya, T.E.~Browder, Q.~Campagna, A.~Datta, S.~Dubey, L.~Mukherjee
  et~al., \emph{{Implications for the \ensuremath{\Delta}AFB anomaly in
  B\textasciimacron{}0\textrightarrow{}D*+\ensuremath{\ell}-\ensuremath{\nu}\textasciimacron{}
  using a new Monte~Carlo event generator}},
  \href{https://doi.org/10.1103/PhysRevD.107.015011}{\emph{Phys. Rev. D}
  {\bfseries 107} (2023) 015011}
  [\href{https://arxiv.org/abs/2206.11283}{{\ttfamily 2206.11283}}].

\bibitem{Shivashankara:2015cta}
S.~Shivashankara, W.~Wu and A.~Datta, \emph{{$\Lambda_b \to \Lambda_c \tau
  \bar{\nu}_{\tau}$ Decay in the Standard Model and with New Physics}},
  \href{https://doi.org/10.1103/PhysRevD.91.115003}{\emph{Phys. Rev. D}
  {\bfseries 91} (2015) 115003}
  [\href{https://arxiv.org/abs/1502.07230}{{\ttfamily 1502.07230}}].

\bibitem{Datta:2017aue}
A.~Datta, S.~Kamali, S.~Meinel and A.~Rashed, \emph{{Phenomenology of $
  {\Lambda}_b\to {\Lambda}_c\tau {\overline{\nu}}_{\tau } $ using lattice QCD
  calculations}}, \href{https://doi.org/10.1007/JHEP08(2017)131}{\emph{JHEP}
  {\bfseries 08} (2017) 131}
  [\href{https://arxiv.org/abs/1702.02243}{{\ttfamily 1702.02243}}].

\bibitem{ParticleDataGroup:2022pth}
{\scshape Particle Data Group} collaboration, \emph{{Review of Particle
  Physics}}, \href{https://doi.org/10.1093/ptep/ptac097}{\emph{PTEP} {\bfseries
  2022} (2022) 083C01}.

\bibitem{FlavourLatticeAveragingGroupFLAG:2021npn}
{\scshape Flavour Lattice Averaging Group (FLAG)} collaboration, \emph{{FLAG
  Review 2021}},
  \href{https://doi.org/10.1140/epjc/s10052-022-10536-1}{\emph{Eur. Phys. J. C}
  {\bfseries 82} (2022) 869}
  [\href{https://arxiv.org/abs/2111.09849}{{\ttfamily 2111.09849}}].

\bibitem{Bolognani:2024cmr}
C.~Bolognani, M.~Reboud, D.~van Dyk and K.K.~Vos, \emph{{Constraining
  $|V_{cs}|$ and physics beyond the Standard Model from exclusive
  (semi)leptonic charm decays}},
  \href{https://arxiv.org/abs/2407.06145}{{\ttfamily 2407.06145}}.

\bibitem{James:1975dr}
F.~James and M.~Roos, \emph{{Minuit: A System for Function Minimization and
  Analysis of the Parameter Errors and Correlations}},
  \href{https://doi.org/10.1016/0010-4655(75)90039-9}{\emph{Comput. Phys.
  Commun.} {\bfseries 10} (1975) 343}.

\bibitem{iminuit}
H.~Dembinski and P.O.~et~al., \emph{scikit-hep/iminuit},  Dec, 2020.
\newblock 10.5281/zenodo.3949207.

\bibitem{Kling:2021gos}
F.~Kling and L.J.~Nevay, \emph{{Forward neutrino fluxes at the LHC}},
  \href{https://doi.org/10.1103/PhysRevD.104.113008}{\emph{Phys. Rev. D}
  {\bfseries 104} (2021) 113008}
  [\href{https://arxiv.org/abs/2105.08270}{{\ttfamily 2105.08270}}].

\bibitem{Abbaneo:2895224}
D.~Abbaneo et~al., \emph{{AdvSND, The Advanced Scattering and NeutrinoDetector
  at High Lumi LHC Letter of Intent}},  Tech. Rep.
  \href{https://cds.cern.ch/record/2895224}{CERN-LHCC-2024-007, LHCC-I-040},
  CERN, Geneva (2024).

\bibitem{Datta:1996gg}
A.~Datta and X.~Zhang, \emph{{Nonuniversal correction to $Z\to b{\bar b}$ and
  single top quark production at Tevatron}},
  \href{https://doi.org/10.1103/PhysRevD.55.R2530}{\emph{Phys. Rev. D}
  {\bfseries 55} (1997) 2530}
  [\href{https://arxiv.org/abs/hep-ph/9611247}{{\ttfamily hep-ph/9611247}}].

\bibitem{Watanabe:2017mip}
R.~Watanabe, \emph{{New Physics effect on $B_c \to J/\psi \tau\bar\nu$ in
  relation to the $R_{D^{(*)}}$ anomaly}},
  \href{https://doi.org/10.1016/j.physletb.2017.11.016}{\emph{Phys. Lett. B}
  {\bfseries 776} (2018) 5} [\href{https://arxiv.org/abs/1709.08644}{{\ttfamily
  1709.08644}}].

\bibitem{Belle:2017rcc}
{\scshape Belle} collaboration, \emph{{Precise determination of the CKM matrix
  element $\left| V_{cb}\right|$ with $\bar B^0 \to D^{*\,+} \, \ell^- \, \bar
  \nu_\ell$ decays with hadronic tagging at Belle}},
  \href{https://arxiv.org/abs/1702.01521}{{\ttfamily 1702.01521}}.

\bibitem{Lubicz:2017syv}
{\scshape ETM} collaboration, \emph{{Scalar and vector form factors of $D \to
  \pi(K) \ell \nu$ decays with $N_f=2+1+1$ twisted fermions}},
  \href{https://doi.org/10.1103/PhysRevD.96.054514}{\emph{Phys. Rev. D}
  {\bfseries 96} (2017) 054514}
  [\href{https://arxiv.org/abs/1706.03017}{{\ttfamily 1706.03017}}].

\bibitem{Azizi:2010zj}
K.~Azizi, R.~Khosravi and F.~Falahati, \emph{{Exclusive $D_{s} \to
  (\eta,\eta^{\prime}) l \nu$ decays in light cone QCD}},
  \href{https://doi.org/10.1088/0954-3899/38/9/095001}{\emph{J. Phys. G}
  {\bfseries 38} (2011) 095001}
  [\href{https://arxiv.org/abs/1011.6046}{{\ttfamily 1011.6046}}].

\bibitem{BaBar:2010vmf}
{\scshape BaBar} collaboration, \emph{{Analysis of the $D^+ \to K^- \pi^+ e^+
  \nu_e$ decay channel}},
  \href{https://doi.org/10.1103/PhysRevD.83.072001}{\emph{Phys. Rev. D}
  {\bfseries 83} (2011) 072001}
  [\href{https://arxiv.org/abs/1012.1810}{{\ttfamily 1012.1810}}].

\bibitem{Fu:2018yin}
H.-B.~Fu, L.~Zeng, R.~L\"u, W.~Cheng and X.-G.~Wu, \emph{{The $D\to \rho$
  semileptonic and radiative decays within the light-cone sum rules}},
  \href{https://doi.org/10.1140/epjc/s10052-020-7758-4}{\emph{Eur. Phys. J. C}
  {\bfseries 80} (2020) 194}
  [\href{https://arxiv.org/abs/1808.06412}{{\ttfamily 1808.06412}}].

\bibitem{Donald:2013pea}
{\scshape HPQCD} collaboration, \emph{{$V_{cs}$ from $D_s \to \phi \ell \nu$
  semileptonic decay and full lattice QCD}},
  \href{https://doi.org/10.1103/PhysRevD.90.074506}{\emph{Phys. Rev. D}
  {\bfseries 90} (2014) 074506}
  [\href{https://arxiv.org/abs/1311.6669}{{\ttfamily 1311.6669}}].

\bibitem{Meinel:2016dqj}
S.~Meinel, \emph{{$\Lambda_c \to \Lambda l^+ \nu_l$ form factors and decay
  rates from lattice QCD with physical quark masses}},
  \href{https://doi.org/10.1103/PhysRevLett.118.082001}{\emph{Phys. Rev. Lett.}
  {\bfseries 118} (2017) 082001}
  [\href{https://arxiv.org/abs/1611.09696}{{\ttfamily 1611.09696}}].

\end{thebibliography}\endgroup

\end{document}